\documentclass[8.5pt,twoside,twocolumn]{article}
\usepackage[super,sort&compress,comma]{natbib} 
\usepackage[version=3]{mhchem}
\usepackage{times,mathptm}
\usepackage{secsty}
\usepackage{balance} 
\usepackage{color}

\usepackage{graphicx} 
\usepackage[format=plain,justification=raggedright,singlelinecheck=false,font=small,labelfont=bf,labelsep=space]{caption} 
\usepackage{fancyhdr}
\begin{document}

\thispagestyle{plain}
\fancypagestyle{plain}{
\renewcommand{\headrulewidth}{1pt}}
\renewcommand{\thefootnote}{\fnsymbol{footnote}}
\renewcommand\footnoterule{\vspace*{1pt}%
\hrule width 3.4in height 0.4pt \vspace*{5pt}} 

\makeatletter 
\renewcommand\@biblabel[1]{#1}            
\renewcommand\@makefntext[1]%
{\noindent\makebox[0pt][r]{\@thefnmark\,}#1}
\makeatother 
\renewcommand{\figurename}{\small{Fig.}~}
\sectionfont{\large}
\subsectionfont{\normalsize} 

\fancyfoot{}
\fancyhead{}
\renewcommand{\headrulewidth}{1pt} 
\renewcommand{\footrulewidth}{1pt}
\setlength{\arrayrulewidth}{1pt}
\setlength{\columnsep}{6.5mm}
\setlength\bibsep{1pt}

\twocolumn[
  \begin{@twocolumnfalse}
\noindent\LARGE{\textbf{Shear dynamics of an inverted nematic emulsion}}
\vspace{0.6cm}

\noindent\large{\textbf{A. Tiribocchi{$^{\ast}$}$^{a}$, M. Da Re{$^{\ast}$}$^{a}$, D. Marenduzzo$^{b}$, E. Orlandini$^{a}$}}

\noindent\textit{\small{\textbf{Received Xth XXXXXXXXXX 20XX, Accepted Xth XXXXXXXXX 20XX\newline
First published on the web Xth XXXXXXXXXX 200X}}}

\noindent \textbf{\small{DOI: 10.1039/b000000x}}
\vspace{0.6cm}


\noindent \normalsize{Here we study theoretically the dynamics of a 2D and a 3D isotropic droplet in a nematic liquid crystal under a shear flow. We find a large repertoire of possible nonequilibrium steady states as a function of the shear rate and of the anchoring of the nematic director field at the droplet surface. 
We first discuss homeotropic anchoring. For weak anchoring, we recover the typical behaviour of a sheared isotropic droplet in a binary fluid, which rotates, stretches and can be broken by the applied flow. For intermediate anchoring, new possibilities arise due to elastic effects in the nematic fluid. We find that in this regime the 2D droplet can tilt and move in the flow, or tumble incessantly at the centre of the channel. For sufficiently strong anchoring, finally, one or both of the topological defects which form close to the surface of the isotropic droplet in equilibrium detach from it and get dragged deep into the nematic state by the flow. In 3D, instead, the Saturn ring associated with normal anchoring disclination line can be deformed and shifted downstream by the flow, but remains always localized in proximity of the droplet, at least for the parameter range we explored.
Tangential anchoring in 2D leads to a different dynamic response, as the boojum defects characteristic of this situation can unbind from the droplet under a weaker shear with respect to the normal anchoring case. 
Our results should stimulate further experiments with inverted liquid crystal emulsions under shear, as most of the predictions can be testable in principle by monitoring the evolution of liquid crystalline orientation patterns or by tracking the position and shape of the droplet over time. 
}
\vspace{0.5cm}
 \end{@twocolumnfalse}
  ]

\footnotetext{\textit{$^{a}$ Dipartimento di Fisica e Astronomia and Sezione INFN di Padova, Universit\'a di Padova, Via Marzolo 8, 35131 Padova, Italy.\\
$^{b}$~SUPA and The School of Physics and Astronomy, University of Edinburgh, Edinburgh EH9 3FD, United Kingdom}}

\section{Introduction}

Dispersions of particles in a host fluid are important examples of soft matter with many potential applications in food industry, drugs, paints and design of new composite materials~\cite{musevic_11}.
While in colloidal suspensions the hosted particles are solid, emulsions are dispersions of liquid droplets coated with a surfactant.
When the particles are dispersed in a nematic liquid crystal, i.e., an anisotropic fluid, where elongated organic molecules are, on average,  aligned  along a common direction (director), additional long-range forces, due to elastic deformations of the director field in proximity of the droplet surface, are induced and topological defects are observed.
Defects mediate anisotropic droplet-droplet interactions of dipolar or quadrupolar type that typically lead to the formation of ordered droplet structures, such as chains and hexagonal lattices~\cite{paul,paul2,Yada_04,Smalyukh_05,Nazarenko_01}.

Due to a large body of theoretical and experimental works, the equilibrium properties of isotropic droplets in a nematic host (inverted nematic emulsions)  are now well understood: depending on the strength and direction of the anchoring of the nematogens at the droplet surface, many equilibrium structures occur, each with a well defined defect and droplet shape conformation~\cite{Nazarenko_01}.
For example, isotropic droplets with normal anchoring of the director field at their surface are often accompanied by a hyperbolic hedgehog defect~\cite{paul,paul2} while, for weaker anchoring
a Saturn ring defect (or a pair of antipodal defects of topological charge -1/2 in 2D) surrounding the droplet is observed~\cite{Mondain_99,Gu_00,Fukuda_01}.
The processes of defects formation and director field orientation in proximity of the droplet surface have been also investigated numerically by different means such as molecular dynamics simulations~\cite{Andrienko_01,Allen_05,Billeter_00,gettelfinger},  Monte Carlo simulations~\cite{Grollau_03,Ruhwandl_97} and through minimisation of a Landau-de Gennes free energy functional~\cite{Stark_99,Zhou_et_al_JFM_2007}, although in many cases the approximation of undeformable droplet has been considered.

Much less is  known, however, on the rheological and  response properties of isotropic-nematic mixtures when they are, for example, subject to external perturbations such as electric, magnetic and flow fields.
In fact, although molecular dynamics approaches are in principle able to describe the time evolution of these systems, they are limited to very small length and short time scales where hydrodynamic modes are still irrelevant.
Continuum models, based on a Landau-de Gennes free energy  description of the nematogens, have been used in the past to study either the phase separation dynamics of symmetric nematic-isotropic mixtures~\cite{Araki_04} or the effect of hydrodynamic flow on the orientational order of the nematic liquid crystal in presence of spherical, rigid, inclusions~\cite{Fukuda_04}.
More recently,  other theoretical studies,  based on a free-energy description of the system and Lattice Boltzmann approaches, have been carried out with the aim of characterising either the equilibrium properties of the interface between nematic and isotropic fluids or the shape of a  droplet of an isotropic fluid immersed in a nematic liquid crystal in presence of a surfactant~\cite{care,lishchuk}.
In particular Sulaiman et al.~\cite{sulaiman}  have recently  introduced and tested a Lattice Boltzmann algorithm which solves numerically  the hydrodynamic equations of motion of a nematic coexisting with an isotropic phase, either in absence or in presence of an electric field.

Here we adapt this algorithm to study the effect that an externally imposed shear flow can have on the dynamical properties of inverted nematic emulsions,  described as a single two- or three-dimensional droplet of isotropic fluid surrounded by a nematic  liquid crystal.
By varying the shear rate and the ratio between the elastic energy scales of the nematic in the bulk and at the droplet surface, we observe a rich  dynamical response. For weak anchoring, the behaviour of the system resembles that of a sheared isotropic droplet in a binary fluid, whereas for intermediate and strong anchoring many other steady states are observed. 
Of particular interest is the occurence of oscillatory steady states in which the droplet tumbles and deforms in the flow (for intermediate anchoring), or in which the topological defects created by the anchoring detach from the droplet surface and move around in the bulk. Such oscillatory steady states are possible since the system is driven far from equilibrium by the applied shear. 
In our work, we characterise the steady states in terms both of the droplet shape (aspect ratio and tilt) and of the defect textures  within the nematic host. 

The paper is structured as follows.
In Section 2, we introduce the model which describes the equilibrium phase behaviour and the hydrodynamics of 2D droplets of isotropic fluid suspended in a lyotropic nematic liquid crystal.
In particular we write down the Landau-de Gennes free energy of the system and the Beris-Edwards equations of motion for nematic liquid crystals coupled to Cahn-Hilliard dynamics for the diffusion of the two species.
In Section 3 we partition our results as follows. First we discuss the equilibrium properties of a single droplet of Newtonian fluid dispersed into a nematic host fluid, for different anchoring conditions (either homeotropic, i.e., normal, or homogeneous, i.e., tangential). We next present the main findings of the paper referring to  the effect that linear shear flow has on the shape and dynamical properties of the emulsion.  The end of the section is devoted to extend the study to 3D systems where the defect configuration at equilibrium  is given by a Saturn ring hugging the droplet. The effect of the shear on the dynamics of the droplet-Saturn ring pair is presented and compared with the 2D counterparts.
Finally Section 4 is devoted to a discussion of the results and to conclusions.

\section{Model and methods}

\subsection{Underlying free energy, and equations of motion}

We consider an inverted nematic emulsion in an extremely diluted regime in which a single isotropic droplet is dispersed in a nematic liquid crystal. 
In this two-component system the droplet
is made up of anisotropic-shaped molecules (such as rods) randomly distributed and oriented in the space.
The resulting phase is that of a liquid crystal in the isotropic phase in which, unlike in the nematic phase, there is neither orientational
nor positional order.
The physics of the system can be 
described in terms of a set of coarse-grained variables $\phi({\vec r},t)$, $\rho({\vec r},t)$,
${\vec v}({\vec r},t)$ and ${\bf Q}(\vec{r},t)$ which are respectively: an order parameter related to 
the relative concentration of the nematic phase ($\phi({\vec r},t)=0$ in the isotropic phase); the mass 
density; the average velocity field and the tensor order parameter that, within the Beris-Edwards theory~\cite{beris}, 
describes the nematic phase. More specifically, in the uniaxial approximation 
$Q_{\alpha\beta}=q_0(\hat{n}_{\alpha}\hat{n}_{\beta}-\frac{1}{3}\delta_{\alpha\beta})$, where $\hat{n}$ is the director field
(Greek subscripts denote Cartesian coordinates)
and $q_0$ is the local degree of nematic order related to the largest eigenvalue of ${\bf Q}$
($0\leq q_0 \leq \frac{2}{3}$).

The equilibrium properties of the system are encoded in a Landau-de Gennes 
free energy ${\cal F}=\int_V fdV$, 
where
\begin{equation}\label{fe}
f=f_{bf}(\phi)+f_{lc}(\phi,{\bf Q})+f_{int}(\phi,{\bf Q})+f_W({\bf Q}).
\end{equation}
The term 
\begin{equation}\label{fe_bf}
f_{bf}(\phi) = \frac{a}{4}\phi^{2}(\phi-\phi_{0})^{2}+\frac{\kappa}{2}\left|\nabla \phi\right|^{2}
\end{equation}
stems from a typical binary fluid formalism and is made by two contributions:  
the first one is a double well potential allowing bulk phase separation into a nematic 
(outside with $\phi\simeq\phi_0$) and isotropic (inside with $\phi=0$) phase in the droplet geometry, whereas 
the second one creates an interfacial tension between these phases whose strength depends on 
$\kappa$.
The term 
\begin{eqnarray}\label{fe_lc}
f_{lc}(\phi,Q)&=& A_0 \biggl[\frac{1}{2}\left(1-\frac{\zeta(\phi)}{3}\right)Q^2_{\alpha\beta}-\frac{\zeta(\phi)}{3}Q_{\alpha\beta}Q_{\beta\gamma}Q_{\gamma\alpha}\nonumber\\
                 &+& \frac{\zeta(\phi)}{4}(Q_{\alpha\beta}^2)^2\biggr] + \frac{K}{2}(\partial_{\alpha}Q_{\beta\gamma})^2,
\end{eqnarray}
is the free energy density of the liquid crystal phase. It is made by four contributions 
(summation over repeated indexes is assumed). The first three, multiplied by the positive constant $A_0$,
are the bulk free energy density for an uniaxial liquid crystal system with an isotropic-nematic transiton 
at $\zeta(\phi)=2.7$. The parameter $\zeta(\phi)$, which determines which phase (isotropic or nematic) 
is the stable one, is assumed to be linearly dependent on the concentration $\phi$, namely
\begin{equation}
\zeta=\zeta_0+\zeta_s\phi,
\end{equation}
where $\zeta_0$ and $\zeta_s$ are constants controlling the boundary of the coexistence region. 
The fourth term creates an elastic penalty for local distortion of the nematic order, within the 
(standard) ``one elastic constant'' approximation~\cite{degennes},  with $K$ being 
 the resulting single elastic constant. 
The term 
\begin{equation}
f_{int}(\phi,Q)=L(\partial_{\alpha}\phi)Q_{\alpha\beta}(\partial_{\beta}\phi)
\end{equation}
takes into accout the anchoring of the nematic liquid crystal on the surface of the 
droplet. The constant $L$ controls the anchoring strength: 
if negative the director is aligned perpendicularly (homeotropic anchoring)
to the surface, whereas if positive the director is aligned tangentially to the surface (planar anchoring). 
Finally, in presence of confining walls, one has to take into account the anchoring of the nematic with these 
boundaries. This is described by the last term of Eq.~(\ref{fe})
\begin{equation}
f_W(Q)=\frac{1}{2}W(Q_{\alpha\beta}-Q_{\alpha\beta}^0)^2,
\end{equation}
where the constant $W$ controls the strength of the nematic anchoring at the walls
and $Q_{\alpha\beta}^0=S_0\left (n^0_{\alpha}n^0_{\beta}-\delta_{\alpha\beta}/3\right )$
with $n^0_{\alpha}$ and $S_0$ being respectively  the direction and the magnitude  of the nematic ordering 
at the walls.

The dynamical equations governing the evolution of the system are

\begin{eqnarray}
\partial_t\phi+\partial_{\alpha}(\phi u_{\alpha})&=&\nabla \left (M \nabla \frac{\delta{\cal F}}{\delta\phi} \right )\label{conc_eq}, \\
(\partial_t+\vec{u}\cdot\nabla){\bf Q}-{\bf S}({\bf W},{\bf Q})&=&\Gamma{\bf H}\label{ord_param_eq},\\
\nabla\cdot{\vec u}&=&0\label{cont_eq},\\
\rho(\partial_t+u_{\beta}\partial_{\beta})u_{\alpha}&=&\partial_{\beta}\sigma_{\alpha\beta}^{total}\label{nav_stokes_eq}.
\end{eqnarray}
The first equation, that governs the time evolution of the concentration $\phi(\vec{r},t)$, 
is a convection-diffusion equation for a model B~\cite{halperin} where $M$ is a thermodynamic mobility parameter and  
$\delta{\cal F}/\delta\phi$ is the chemical potential.

The dynamics of the liquid crystal order parameter, the tensor ${\bf Q}$, is described by Eq.~(\ref{ord_param_eq}) 
which is a convection relaxation equation. 
The first two terms on the left hand side are the material derivative. Moreover, 
since for rod-like molecules the order parameter distribution can be rotated and stretched by flow gradients,
a further contribution ${\bf S}({\bf W},{\bf Q})$  is needed~\cite{beris}.
Its explicit expression is 
\begin{eqnarray}
{\bf S}({\bf W},{\bf Q})&=&(\xi{\bf D}+{\bf \omega})({\bf Q}+{\bf I}/3)+({\bf Q}+{\bf I}/3)(\xi{\bf D}-{\bf \omega})\nonumber\\
&-&2\xi({\bf Q}+{\bf I}/3)Tr({\bf Q}{\bf W}),
\end{eqnarray}
where ${\bf D}=({\bf W}+{\bf W}^T)/2$ and ${\bf \omega}=({\bf W}-{\bf W}^T)/2$ are the symmetric and antisymmetric parts, respectively,
of the velocity gradient tensor $W_{\alpha\beta}=\partial_{\beta}u_{\alpha}$ and ${\bf I}$ is the unit matrix.
The constant $\xi$ takes into account the aspect ratio of the molecules of a given liquid crystal and determines the dynamical behaviour
of the director field under shear, in particular whether it is  flow aligning ($\xi\geq 0.6$), giving a stable response,  or flow tumbling ($\xi < 0.6$),
generating an unsteady response.
On the right hand side $\Gamma$ is the collective rotational diffusion constant and 
\begin{equation}
{\bf H}=-\frac{\delta{\cal F}}{\delta{\bf Q}}+\frac{{\bf I}}{3}Tr\frac{\delta{\cal F}}{\delta{\bf Q}}.
\end{equation}
is the molecular field (this is the analogue for ${\bf Q}$ of the chemical potential $\delta{\cal F}/\delta\phi$ ).

Finally, the last two equations are respectively the continuity and the Navier-Stokes equations for an incompressible fluid.
Here $\sigma^{total}$ is the total hydrodynamic stress which is the sum of three contributions. The first term  is the
viscous stress 
\begin{equation}
\sigma^{visc}_{\alpha\beta}=\eta(\partial_{\alpha}u_{\beta}+\partial_{\beta}u_{\alpha}),  
\end{equation}
where $\eta$ is an isotropic shear viscosity~\cite{beris}. The second one is  the contribution to the stress due to the liquid 
crystalline order and is given by
\begin{eqnarray}
\sigma^{lc}_{\alpha\beta}&=&-P\delta_{\alpha\beta}-\xi H_{\alpha\gamma}(Q_{\gamma\beta}+\frac{1}{3}\delta_{\gamma\beta})-\xi(Q_{\alpha\gamma}+\frac{1}{3}\delta_{\alpha\gamma})H_{\gamma\beta}\nonumber\\&+&
2\xi(Q_{\alpha\beta}-\frac{1}{3}\delta_{\alpha\beta})Q_{\gamma\mu}H_{\gamma\mu}+Q_{\alpha\nu}H_{\nu\beta}-H_{\alpha\nu}Q_{\nu\beta},\nonumber\\
\end{eqnarray}
where the pressure $P$ is given by $P=\rho T -\frac{K}{2}(\nabla{\bf Q})^2$ (where $T$ is the temperature),
and, except in proximity of the droplet surface and of the defect cores, is costant in our simulations to a very good approximation. The last term
is the sum of the interfacial stress between the isotropic and the liquid crystal phases with the elastic stress due to 
the distortions within  the liquid crystal phase
\begin{equation}
\sigma_{\alpha\beta}^{s}=-\left(\frac{\delta{\cal F}}{\delta\phi}\phi-{\cal F}\right)\delta_{\alpha\beta}-\frac{\delta{\cal F}}{\delta(\partial_{\beta}\phi)}\partial_{\alpha}\phi-
\frac{\delta{\cal F}}{\delta(\partial_{\beta}Q_{\gamma\mu})}\partial_{\alpha}Q_{\gamma\mu}.
\end{equation}

\subsection{Numerical aspects and mapping to physical units}

Eqs.~(\ref{conc_eq},\ref{ord_param_eq},\ref{cont_eq},\ref{nav_stokes_eq}) are solved numerically by using a hybrid lattice Boltzmann method previously 
used in similar systems such as binary fluids~\cite{tiribocchi1}, liquid crystals~\cite{tiribocchi2,prl} 
and active matter~\cite{cates1,elsen,elsen2}. 
Unless explicitely stated otherwise, most of the simulations are performed 
on a two-dimensional rectangular lattice ($Lx=400$, $Ly=120$) in which an isotropic droplet is initially placed 
at its centre and surrounded by a nematic liquid crystal. 
The entire system is sandwiched  between two parallel flat walls. 
We choose a rectangular box with a longer size along the shear direction 
in order to  minimize the possible effects of interference between 
periodic images of the droplet moving with  the flow. 

The concentration field $\phi$ is initially set to zero inside the droplet and to a constant value $\phi_0$ 
in the bulk nematic phase. 
Similarly, the order parameter ${\bf Q}$ is initially set to zero inside the droplet and different from zero elsewhere.
In particular the initial direction of the director in the nematic phase equals the one imposed at the  walls.
This means that the director in the bulk is along the $y$-direction for homeotropic (or perpendicular) anchoring at 
walls  and along the $x$-direction if homogeneous (or tangential) anchoring is instead considered. 
We incidentally note that these initial conditions lead to a droplet whose final size, after equilibration,
depends on the anchoring conditions of the director on its surface. 
A similar effect has been observed in Ref.~\cite{giulio} in which an isotropic droplet is embedded in a 
polar liquid crystal,
and has been ascribed to the deviation of the equilibrium values of $\phi$ of the two phases from the ones at the minima of 
the free energy (which are $\phi=0,2$). This deviation however is,  in terms of the radius of the droplet,  $<1\%$ 
and does not affect in an appreciable way the dynamics of the droplet under shear.
Finally at the walls we impose  no-slip boundary conditions for the velocity field,  
neutral-wetting (meaning that there are no flows of matter across the walls~\cite{tiribocchi1}) for the concentration field and
strong anchoring of the nematic  (this is achieved by setting the anchoring strength $W=0.4$).

Starting from this initial set up, the system is let to relax into its equilibrium state that 
afterwards we shear by moving both walls along opposite directions at constant speed. 
In most simulations the following parameter values have been used: $a=7\times 10^{-2}$, $D=5\times 10^{-2}$, $\phi_0=2$, $\Gamma=1$, $A_0=1$
and $\eta\simeq 1.67$. 
Note that the parameter $\xi$, appearing in Eq.~(\ref{ord_param_eq}), tunes the dynamical response of the 
director field under shear. We have set $\xi=0.7$ which is a value within the flow aligning regime where,
in a linear velocity profile, the director field reaches a steady state characterised by a given
flow aligning angle (the Leslie's angle)~\cite{degennes}. 
For completness we have also perfomed simulations with $\xi=0.5$. This  corresponds to the flow tumbling regime 
where a steady solution under a constant shear rate no longer exists and the director rotates in a direction consistent 
with the vorticity of the flow. Although the overall dynamics of the director field in both regimes is 
diverse, we have not found appreciable differences in the defect dynamics under a shear flow and here 
we only report the flow aligning case.

The key parameters for determining the shape of the droplet are the tension coefficient, which we have kept 
fixed to $\kappa=0.14$ (for stability reasons), the interface cross-gradient coefficient $L$, which ranges from $\pm 10^{-2}$ to $\pm 6\times 10^{-2}$ 
(negative values for strong homeotropic anchoring and positive values for strong tangential anchoring), and the
elastic constant $K$ whose values varies between  $8\times 10^{-3}$ and $8\times 10^{-2}$. 
A suitable dimensionless quantity which characterises the shape of the droplet is the ratio
$\lambda ={\cal F}_{int}R/(\Sigma K)$, where ${\cal F}_{int}=\int_{V}f_{int}dV$
(measured in units of $N=J m^{-1}$), $\Sigma$ is the perimeter of the droplet 
(see Appendix 1 for the calculation of $\Sigma$ when the droplet shape is deformed under shear) and $R$ its radius. 
This quantity measures the strength of the surface anchoring relative to the bulk elastic deformations~\cite{degennes}. Typical values for the anchoring strength ${\cal F}_{int}/\Sigma$ found in the literature~\cite{cognard,anderson} range from $10^{-7}$ J m$^{-2}$ to $10^{-3}$ J m$^{-2}$, for an elastic constant $K$ of $10^{-11}$N and for a droplet size of $1$-$10\mu$m. 
This leads to a corresponding value of $\lambda$ varying within the interval $10^{-2}$-$10^{3}$. 
If $\lambda < 1$  surface anchoring is weak compared to bulk nematic distortions, whereas if $\lambda \gg 1$ it is larger and can significavely deform the director near the droplet.
A further quantity which affects the droplet shape is the surface tension $\sigma={\cal F}_{{\kappa}_{\phi}}/\Sigma$, 
where ${\cal F}_{{\kappa}_{\phi}}=\int_V dV{\kappa}/2(\nabla\phi)^2$.
If $\kappa=0.14$, $\sigma\simeq 10^{-3}$ in simulation units. This corresponds to a physical value of $10^{-5}$ J m$^{-2}$,
whereas experimental values vary in the range 
$\simeq (10^{-5}-10^{-2})$ J m$^{-2}$~\cite{faetti,cognard,anderson}.
Unless stated otherwise, we have kept this quantity constant.
In the next section we show that, depending on the values of these parameters, several possible equilibrium 
shapes, whose structure is affected by the position of topological defects, can be identified.
These equilibrium states will in turn respond differently when sheared.
Similar shapes have been found and  discussed in Ref.~\cite{sulaiman}. 

Note that all the aforementioned  values are in simulation units. 
In order to map them into physical units we follow the approach given in  Ref.~\cite{sulaiman}.
In particular, assuming to model a flow-aligning regime, $\sim \mu$m thick device with a rotational 
viscosity of roughly $1$ poise (a typical value of 5CB), one gets
the length-scale and time-scale to be respectively $\Delta x=10^{-7}$m and $\Delta t=10^{-6}$s. Furthermore, 
an elastic constant $K$ ranging between $8\times 10^{-3}-8\times 10^{-2}$ in simulation units, corresponds to $\simeq 10-100$ pN, within
the typical values of nematic liquid crystals. To compare the effect of anchoring in simulations and experiments, we can use the dimensionless parameter $\lambda$ defined previously. 

\section{Results and discussion}

We first characterize the equilibrium properties of an isotropic droplet in nematic  when either homogeneous or homeotropic anchoring of the director at the surface of the droplet is considered. For each case we also look at the effect that different directions of the anchoring at the walls (i.e. either perpendicular or parallel) may have on the equilibrium configurations.

Later on the dynamical response of the equilibrated configurations when subject to a shear flow (flow aligning regime) 
is studied for different shear rates.

\subsection{Isotropic droplet in a nematic phase at equilibrium}

\begin{figure*}[htbp]
\centerline{\includegraphics[width=0.92\textwidth]{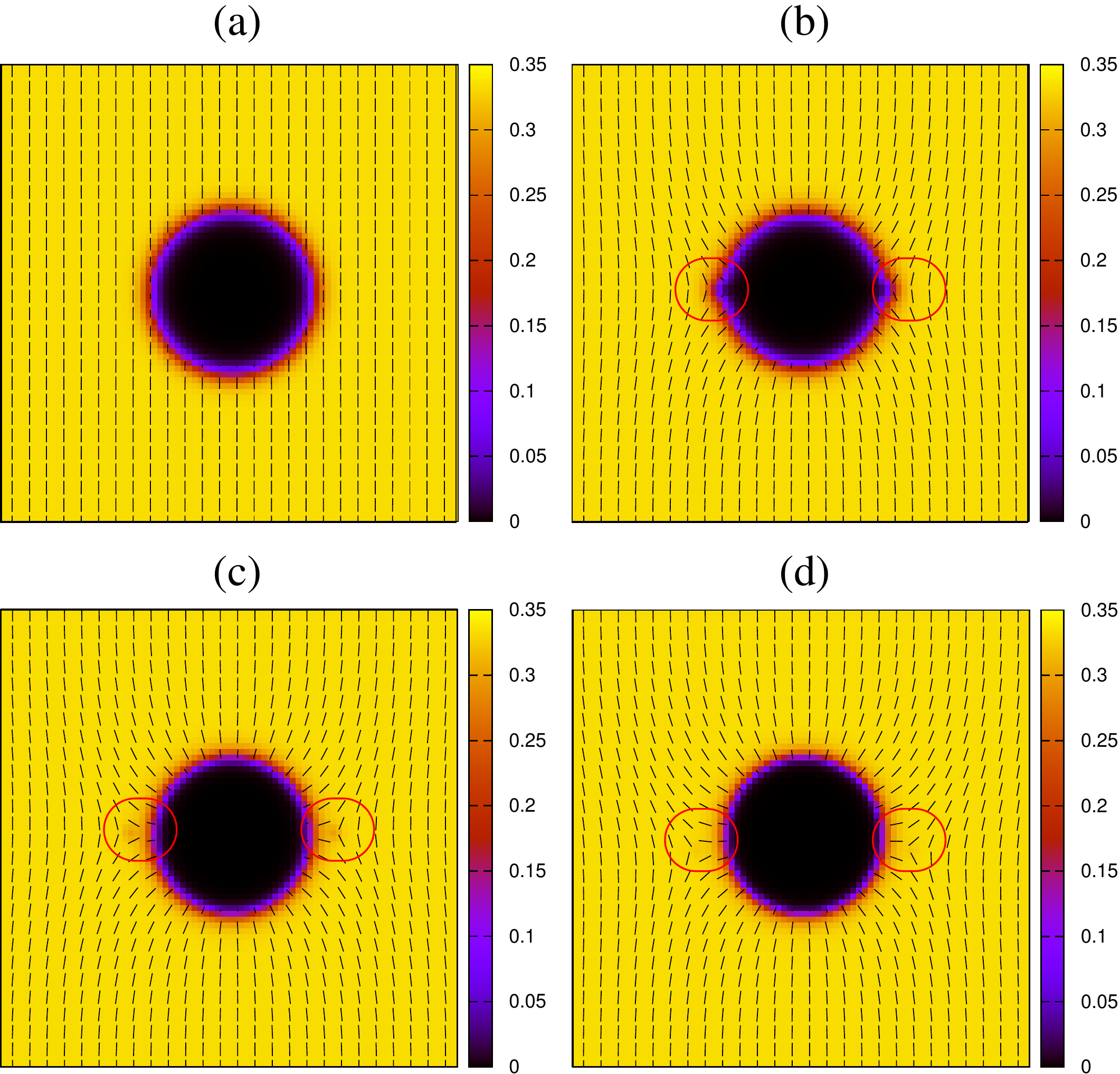}}
\caption{Equilibrium director profile of an isotropic droplet in a nematic host when homeotropic anchoring is set on 
its surface, while on both walls there is strong homeotropic anchoring.
In the background is shown the corresponding profile of the largest eigenvalue of the tensor order parameter ${\bf Q}$:
this goes from zero, inside the droplet (black region), to $\simeq 0.33$ in the nematic
phase (yellow region). Red circles indicate the position of the defects.
Parameters are: (a) $K=8\times 10^{-2}$, $L=-4\times 10^{-4}$, $\lambda\simeq 2.8\times 10^{-3}$; (b) $K=8\times 10^{-2}$,  $L=-4\times 10^{-2}$,
$\lambda\simeq 1.1$; (c) $K=2.5\times 10^{-2}$, $L=-4\times 10^{-2}$, $\lambda\simeq 4.1 $; (d) $K=8\times 10^{-3}$, $L=-4\times 10^{-2}$, $\lambda\simeq 13$.
As mentioned in the text the simulation box is rectangular but here only its central part is shown.}
\label{fig1}
\end{figure*}

We first consider an isotropic droplet of radius $R=15$ located at the centre of a rectangular box of 
size $L_x=400$, $L_y=80$. The anchoring on the surface of the droplet is 
homeotropic and the director is perpendicularly anchored on both walls. 
The corresponding simulation is run for $5\times 10^5$ time-steps until the droplet and the surrounding medium are completely equilibrated, a state achieved when the total free energy is at its minimum. 
In Fig.~\ref{fig1} we show four equilibrium configurations obtained by appropriately changing the elastic constant $K$ 
and the surface anchoring strength $L$. In terms of the number $\lambda$, these correspond to 
$\lambda\simeq 2.8\times 10^{-3}$ (Fig.~1a), $\lambda\simeq 1.1$ (Fig.~1b), $\lambda\simeq 4.1$ (Fig.~1c), $\lambda\simeq 13$ 
(Fig.~1d)\footnote{Since ${\cal F}_{int}$ is negative we will implicitly consider its absolute value.}.
When strong homeotropic anchoring is set on the surface of the droplet (see Fig.~\ref{fig1}), an imaginary defect of topological charge $1$ is enucleated at its centre.
The conservation of the topological charge requires though the formation of two defects of topological charge $-1/2$, which are located on opposite parts of the droplet
(see, for instance, Fig.~\ref{fig1}c or d). This is the 2D version of the  well-known Saturn ring~\cite{paul,paul2,luben}\footnote{Other configurations are possible, such as a hyperbolic hedgehog (a defect of charge
$-1$ located on one side of the droplet) or a disclination ring of a finite radius (located above or below the droplet). An extensive discussion can be found in Ref.~\cite{luben}}. 
The position of the defects pair can be controlled by properly balancing the strength of the surface anchoring and the bulk elastic distortions (namely the parameter $\lambda$). 
For $\lambda\ll 1$, for instance, surface anchoring is very weak and defects disappear, leaving the shape of the droplet unaltered (Fig.~\ref{fig1}a). Notice that this is in agreement with the 
topological charge conservation, as in absence of anchoring the imaginary defect (of charge $1$) does not form inside the droplet, maintaining the total charge zero.
When the anchoring strength becomes comparable with the elastic nematic energy ($\lambda\simeq 1$), defects appear on the surface where they are firmly anchored (Fig.~\ref{fig1}b), 
favouring a nutshell-like shape of the droplet. For higher values of $\lambda$, both defects emerge on opposite sides of the droplet along the equator (Fig.~\ref{fig1}c).
Lastly, for $\lambda\gg 1$, since the elastic liquid crystal energy becomes very small compared to the surface energy,  
the defects pair are clearly  far apart from  the droplet and well inside the bulk nematic phase. 
Similar droplet configurations have been also discussed in Ref.~\cite{sulaiman} in which their dynamics in presence of an applied electric field is studied, and in Ref.~\cite{lishchuk}  where the role of a surfactant has also been taken into account.

\begin{figure*}[htbp]
\centerline{\includegraphics[width=0.92\textwidth]{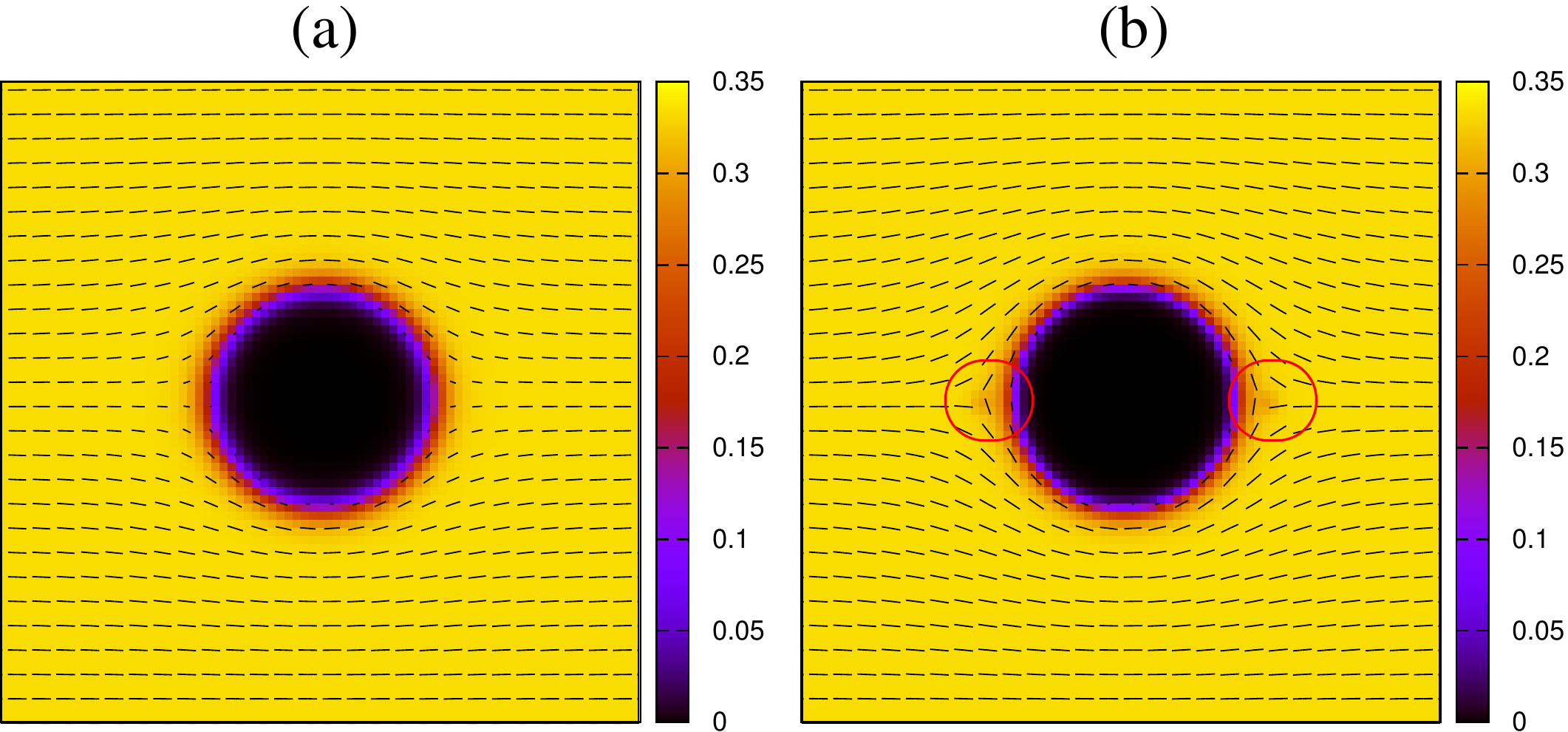}}
\caption{Equilibrium director profile of an isotropic droplet in a nematic host when tangential anchoring is set on its surface. The anchoring is strong and homogeneous on both walls.
On the background the corresponding profile of the largest eigenvalue of the tensor order paramter ${\bf Q}$, which goes from zero inside the droplet (black region) to $\simeq 0.33$ in the nematic 
liquid crystal phase (yellow region). Red circles indicate the position of the defects. Paramters are: (a) $K=8\times 10^{-2}$, $L=4\times 10^{-2}$, $\lambda\simeq 0.56$ and (b) $K=2.5\times 10^{-2}$, $
L=4\times 10^{-2}$, $\lambda\simeq 1.92$. Only the central part of the rectangular box is shown.}
\label{fig2}
\end{figure*}

For homogeneous anchoring at the droplet surface, different steady states are expected. 
In this case we have considered an isotropic droplet of radius $R=15$
embedded in the nematic phase with homogeneous anchoring on both walls. 
This choice determines a final configuration in which two defects of topological charge $-1/2$
are located on opposite sides along the equator, similarly to what obtained for the states of Fig.~\ref{fig1}. 
By varying the values of the nematic elastic constant $K$ we have identified two equilibrium shapes (see Fig.~\ref{fig2}). 
When the distortion energy overcomes the surface one, defects are completely absorbed within the droplet and the director profile smoothly surrounds its surface (Fig.~\ref{fig2}a). 
When both effects become comparable, two defects 
form on both side of the droplet, keeping unaltered the total topological charge (Fig.~\ref{fig2}b). 
The position of defects changes if homeotropic anchoring is set on both walls.
In this case they will form along the north-south direction of the system, on opposite sides of the droplet (not shown).
However this final state is unstable to small perturbations and both defects eventually shift and 
move along the droplet surface until they find a more stable conformation. 

We finally mention that, in line with previous nematodynamics studies~\cite{sluckin,callan}, we found a small degree of biaxiality especially
close to defects. By following the approach used in Ref.~\cite{callan}, regions of biaxial order can be found by calculating the values of 
three parameters, namely $e_l=\theta_1-\theta_2$, $e_p=2(\theta_2-\theta_3)$ and 
$e_s=3\theta_3$, where $\theta_1$, $\theta_2$ and $\theta_3$ (with $\theta_1\ge \theta_2\ge \theta_3$) are the eigenvalues of the diagonalised matrix $D_{\alpha\beta}=Q_{\alpha\beta}+\delta_{\alpha\beta}/3$. These parameters
have the following properties: $0\leq e_l, e_p, e_s \leq 1$ and $e_l+e_p+e_s=1$.  
A well-ordered uniaxial nematic arrangement will give $e_l\simeq 1$, whereas regions of isotropy
and of planar ordering (the biaxial state) correspond 
to $e_s\simeq 1$ and $e_p\simeq 1$, respectively. 
In particular we found $e_p\simeq 0.008$ near the defects and $e_p\simeq 0$
far from them, either in the nematic or in the isotropic phase.

\subsection{Isotropic droplet in a nematic phase under shear}

By starting from the equilibrated droplets previously described, we now impose a shear flow on the system 
by moving the top wall along the $y$-axis with velocity $u$ and the bottom wall in the opposite direction with velocity $-u$. 
This sets a  shear rate $\dot{\gamma}=2u/L_y$ measured in $\Delta t^{-1}$ in simulation units. 
In all the cases studied, for low shear rates (typically for  $\dot{\gamma}\simeq 10^{-4}$) the initial droplet configuration is the same as for an unsheared system,
independently on the values of the elastic constant $K$ and on the surface anchoring constant $L$.  We will therefore discuss only results whose shear rate is strong
enough to induce deformations and/or droplet motion.

Besides the parameter $\lambda$, the shape of the droplet and the role played by the forces in the system can be characterized by introducing the capillary number 
$Ca=\frac{R{\dot{\gamma}}\eta\Sigma}{{\cal F}_{{\kappa}_{\phi}}}$. This adimensional quantity, often used in rheological experiments,  measures the strength of viscous forces relative to the surface 
tension acting 
at the interface between two immiscible fluids.  In these simulations $Ca$
is expected to range from $\sim 0.1$ to $\sim 1$. 
To quantify the effect of the shear on the droplet shape we also consider the parameter $D=\frac{a-b}{a+b}$~\cite{taylor} 
($a$ and $b$ represent the major and the minor axis respectively), 
which measures the droplet deformation under shear and goes from $0$ (no deformation) to $1$ (full deformation). A further parameter worth considering is the Reynolds number $Re=\rho\dot{\gamma}R^2/\eta$, 
which measures the importance of inertial forces relative to the viscous ones: for low shear $Re \ll 1$ but, 
as the shear rate increases, $Re$ increases up to $\sim 1$ where inertial forces become comparable 
with the viscous ones and the condition of laminar flow is less complied with.

\begin{figure*}[htbp]
\centerline{\includegraphics[width=0.92\textwidth]{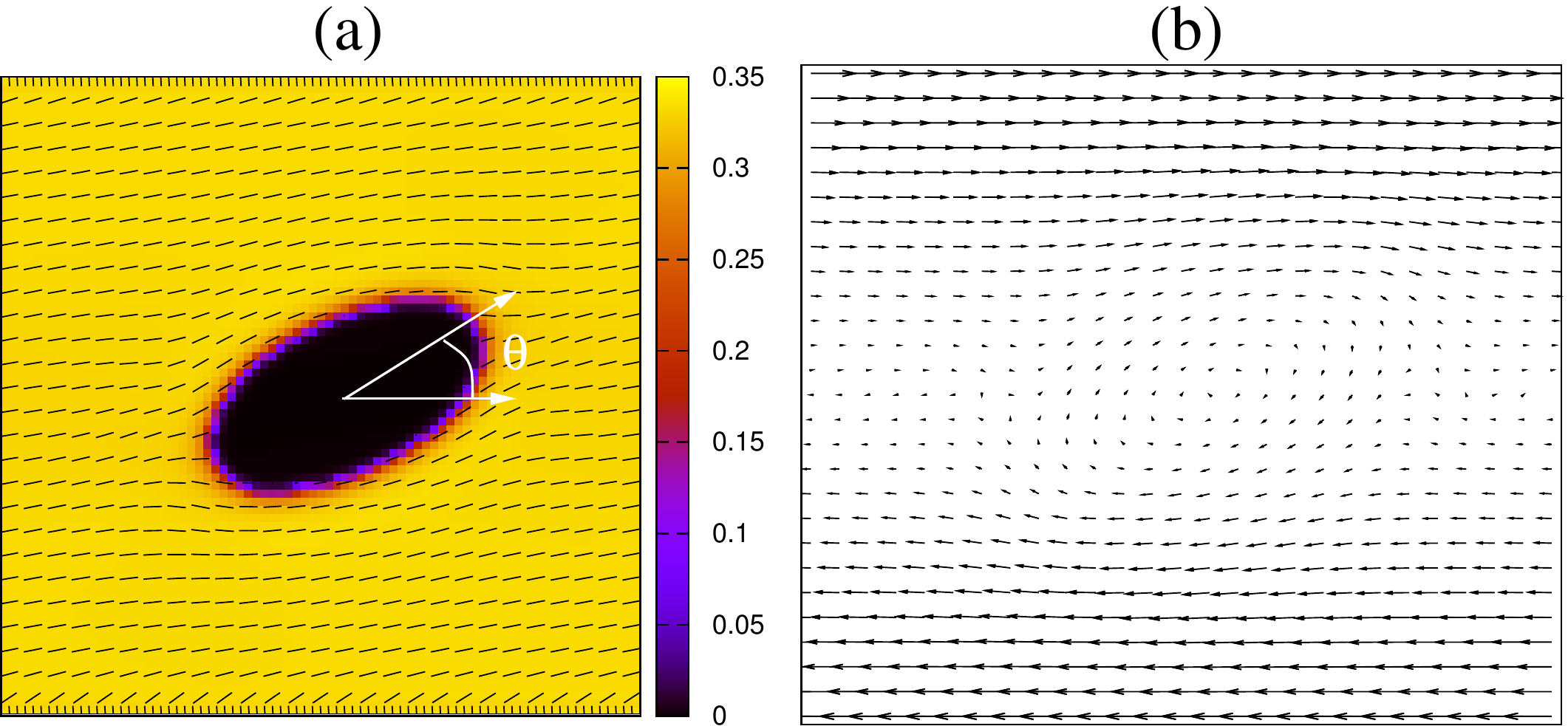}}
\caption{(a) Steady state director profile of an isotropic droplet in a nematic host with no surface anchoring ($L=0$). The anchoring is strong and homeotropic on both walls.
The angle $\theta$ indicates the direction of the major axis of the droplet with the shear direction.
The elastic constant is $K=8\times 10^{-3}$. (b) Steady state velocity field when a shear is applied. A vortex forms inside the droplet whereas intense flow fields, in opposite directions, appear near the walls.}
\label{fig3}
\end{figure*}

A well-known result in binary fluid dynamics is that, if an immiscible isotropic emulsion is dispersed in a Newtonian fluid, one would expect that, under shear, the droplet deforms and orients along the shear flow. Therefore, a first check to validate our model consists in simulating a sheared system in which an isotropic droplet with no surface anchoring ($L=0$, or $\lambda=0$) is embedded into a nematic liquid crystal having a modest or low elastic energy. This is the closest approximation of our system to the simple binary fluid case. 
In Fig.~\ref{fig3}a we show the steady state attained by the droplet after imposing a shear (with $\dot{\gamma}=1.25\times 10^{-3}$) and the director profile of the nematic phase. 
Similarly to what observed in immiscible mixtures of Newtonian fluids, the droplet elongates and afterwards aligns along the direction of the shear flow, with a deformation rate that, at steady state, is  $D \simeq 0.313$. 
The velocity field (shown in Fig.~\ref{fig3}b) is more intense near the walls and weak in the central region of the lattice, with a clockwise vortex generated by the droplet.
The angle $\theta$ formed by the major axis of the droplet with the shear direction is $\simeq 25.7$ degrees, very close to values reported in literature for a single isotropic phase droplet under shear at low Reynolds number (for instance in Ref.~\cite{renardy} the authors found an angle of $25$ degrees for $Re=1$). 
Note that in our case  $Re\simeq 0.31$ and $Ca\simeq 0.663$. These numbers indicate that inertial forces are still weak and the whole system operates in the Stokes regime.
In addition, the director field aligns along the shear flow far from the droplet and in its neighbourhood follows almost everywhere the direction of the major axis.
These results will be assumed as benchmark to be compared with the other cases discussed in the following sections, in which the surface anchoring is no longer neglected (meaning $\lambda > 0$) and higher shear rates are considered.

Finally, we mention a couple of subtle points which need to be kept 
in mind in setting and interpreting our simulations. 
First, it is well known that, by increasing the shear-rate, the temperature at which 
the isotropic-nematic transition occurs decreases~\cite{olmsted}. This would correspond to 
a conversion of the isotropic phase of the droplet into a nematic one. In our simulations we avoid this effect 
by keeping the shear rate small enough.   
Second, one needs to avoid interface-interface interactions that can occur in strongly deformed droplets 
under strong shear. This is achieved by considering a droplet with a sufficiently large radius (provided that its interaction with walls at equilibrium remains negligible) embedded in a relatively long rectangular lattice (necessary to diminish the effect of the periodic image of the droplet). 
In our simulations, as the typical interface thickness separating the droplet and  the liquid crystal is around $8$ lattice sites (which corresponds to roughly $1 \mu$m), a radius of at least $15$ lattice sites is needed.
In the next Sections we will show the results obtained for droplets with a larger radius ($R=22$) and compare these with the ones observed for a smaller radius ($R=15$). In the former, the equilibrated droplet states are the same as the ones seen for the smaller radius even though obtained for slightly different values of $\lambda$. 

\subsubsection{Weak homeotropic anchoring}

We initially consider the case in which homeotropic anchoring is set on the surface of a droplet. 
This anchoring can be experimentally achieved by means of chemical treatments, such as coating the droplet with a surfactant which favours a perpendicular alignment.
As previously mentioned, we have simulated a droplet of radius $R=22$ located in a rectangular box of size $Lx=400$, $Ly=120$ (area fraction at equilibrium
$A_{iso}^{eq}/A_{lc}^{eq}\simeq 3.3\times 10^{-2}$) and a droplet of radius $R=15$ in a box of size $Lx=400$, $Ly=80$ (area fraction at equilibrium
$A_{iso}^{eq}/A_{lc}^{eq}\simeq 2.2\times 10^{-2}$). 

When $\lambda\ll 1$ the equilibrated droplet assumes the shape of Fig.~\ref{fig1}a (independently of the radius). The dynamics under a moderate shear (namely 
for $10^{-4}<\dot{\gamma}<5\times 10^{-4}$) is expected to be very similar to the case discussed in Fig.~\ref{fig3}, in which a droplet with no surface anchoring ($L=0$) has been studied. 
Indeed for $R=22$, besides the different elastic constant (now $K=5\times 10^{-2}$) and the weak contribution of the surface anchoring ($L=-10^{-2}$), the shear stretches and elongates the droplet along the shear flow without generating any net motion 
\footnote{We set $\dot{\gamma}=1.67\times 10^{-4}$ necessary to avoid an excessive shrinkage of the isotropic phase (droplet), hence $Re\simeq 0.1$ at the steady state.}.
In particular the droplet reaches a steady state with the major axis forming an angle $\theta\simeq 43$ degrees with the shear direction. 
The steady state values of the deformation and capillary number are respectively  $D\simeq 0.095$ and $Ca\simeq 0.129$, not very far from the ones obtained for the $L=0$ case. A very similar dynamical behaviour is observed for $R=15$. In this case an angle $\theta\simeq 40.5$ degrees is attained at the steady state
when $\dot{\gamma}=2.5\times 10^{-4}$, with a deformation rate $D\simeq 0.082$ and a capillary number $Ca\simeq 0.12$

\begin{figure*}[htbp]
\centerline{\includegraphics[width=0.92\textwidth]{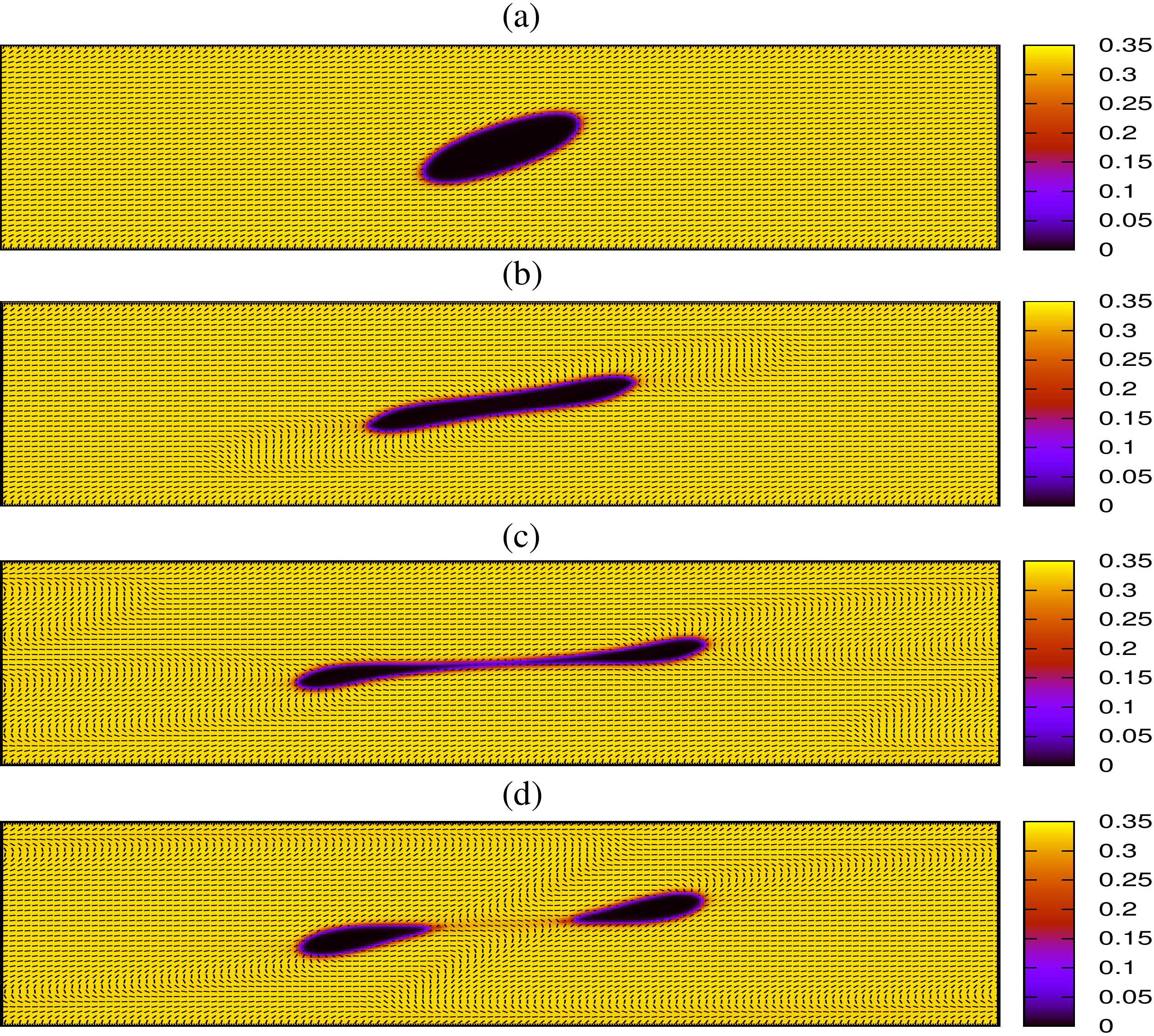}}
\caption{
Dynamic evolution of a droplet of radius $R=22$ without defects under shear with $\dot{\gamma}=1.25\times 10^{-3}$, $K=5\times 10^{-2}$ and $L=-10^{-2}$ (at equilibrium without shear $\lambda\simeq 0.49$). The droplet is initially stretched by the fluid (a) and then undergoes a deep deformation which, after an intermediate rod-like state (b), breaks (c) and generates two separate droplets moving along opposite directions (d).}
\label{fig4}
\end{figure*}

An increase of the shear rate determines a shrinkage of the droplet (regardless of its size) and may lead to its disappearance for very high $\dot{\gamma}$. However larger droplets permit the study of the physics in a wider dynamical range and unveil unexpected properties. For instance, a well-know effect observed on a droplet in a binary fluid mixture is its break-up as the shear is increased~\cite{Wagner}. This phenomenon is still observed when isotropic droplets are surrounded by a nematic liquid crystal, regardless of the presence of topological defects. 
In particular, we have identified two possible rupture regimes related to two different values of $\dot{\gamma}$. For $\dot{\gamma}=1.25\times 10^{-3}$ the droplet breaks in two smaller droplets whereas for higher values, for instance $\dot{\gamma}=1.67\times 10^{-3}$, three smaller droplets result. In Fig.~\ref{fig4} we show the dynamics of the former case.
The droplet, initially moderately stretched along the shear direction, afterwards undergoes a deeper deformation (see Fig.~\ref{fig4}a-b) which further elongates it. Due to the high shear rate, the rod-like state exhibited in Fig.~\ref{fig4}b is however unstable, and the droplet breaks (Fig.~\ref{fig4}c) and divides in two separate smaller droplets (Fig.~\ref{fig4}d) which move along opposite directions. Before the rupture the rate of deformation increases up to $D\simeq 0.9$.
Unlike the single droplet case in which the velocity field displays one vortex inside the droplet (as in Fig.~\ref{fig3}b), now two vortices are formed in the two smaller droplets, with a magnitude  much lower than the one near the walls. At late times both resulting droplets gradually shrink and completely disappear.

\subsubsection{Intermediate and strong homeotropic anchoring: the role of defects}
\begin{figure*}[htbp]
\centerline{\includegraphics[width=0.92\textwidth]{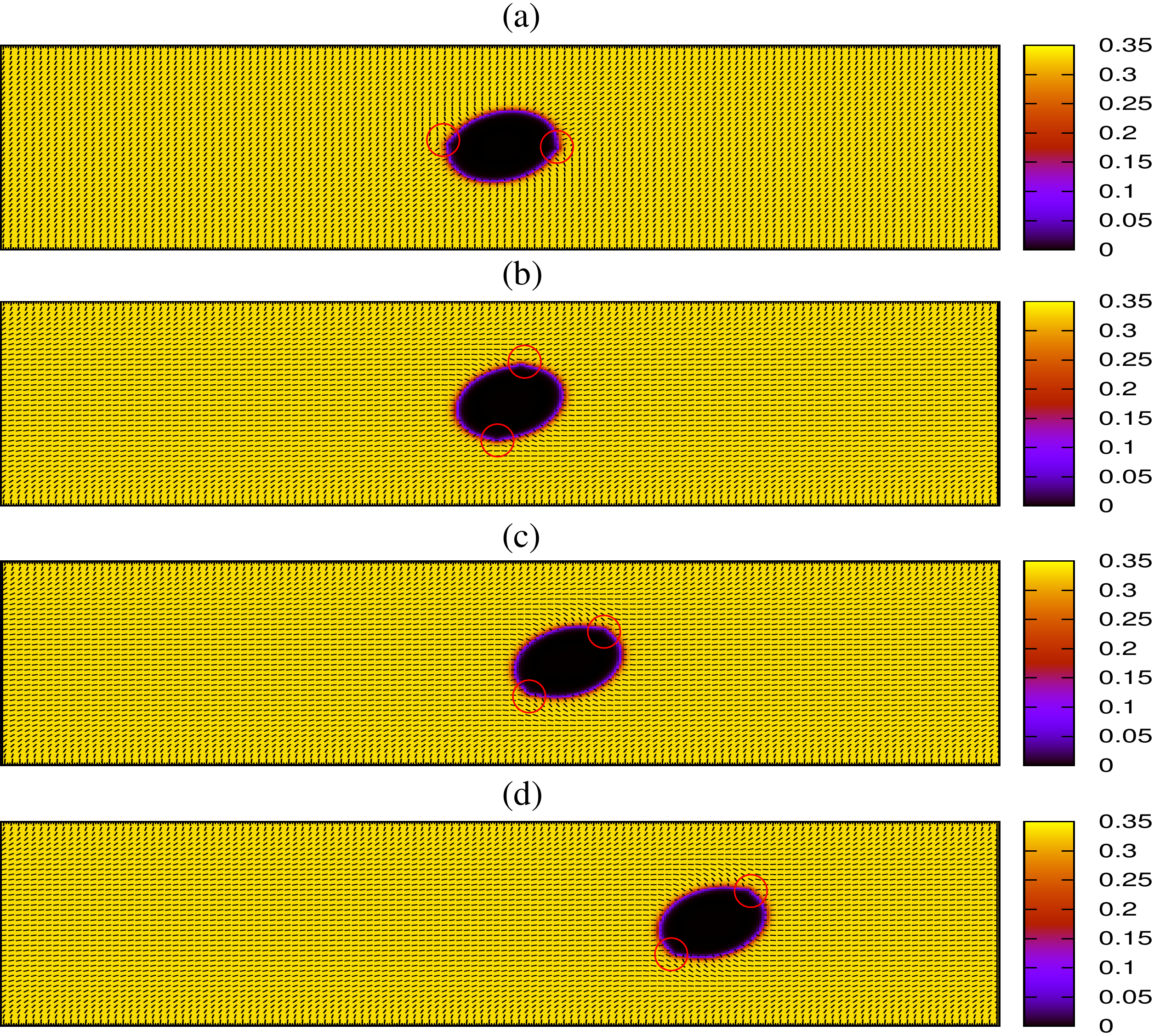}}
\caption{Dynamic evolution of the droplet of radius $R=22$ with defects located on its surface (see Fig.~\ref{fig1}b for the corresponding equilibrium state)  
with $\dot{\gamma}=1.67\times 10^{-4}$. While they rotate in the clockwise direction ((a)-(b)), the droplet deforms and elongates along the shear flow ((a)-(b))
and, at the same time, acquires unidirectional motion along the $x$-axis dragged by the fluid ((c)-(d)).
The capillary number is $\simeq 0.127$ at the steady state and $A_{iso}/A_{lc}\simeq 3.3\times 10^{-2}$. See Movie S1 for the complete dynamics.}
\label{fig5}
\end{figure*}

\begin{figure*}[htbp]
\centerline{\includegraphics[width=1.05\textwidth]{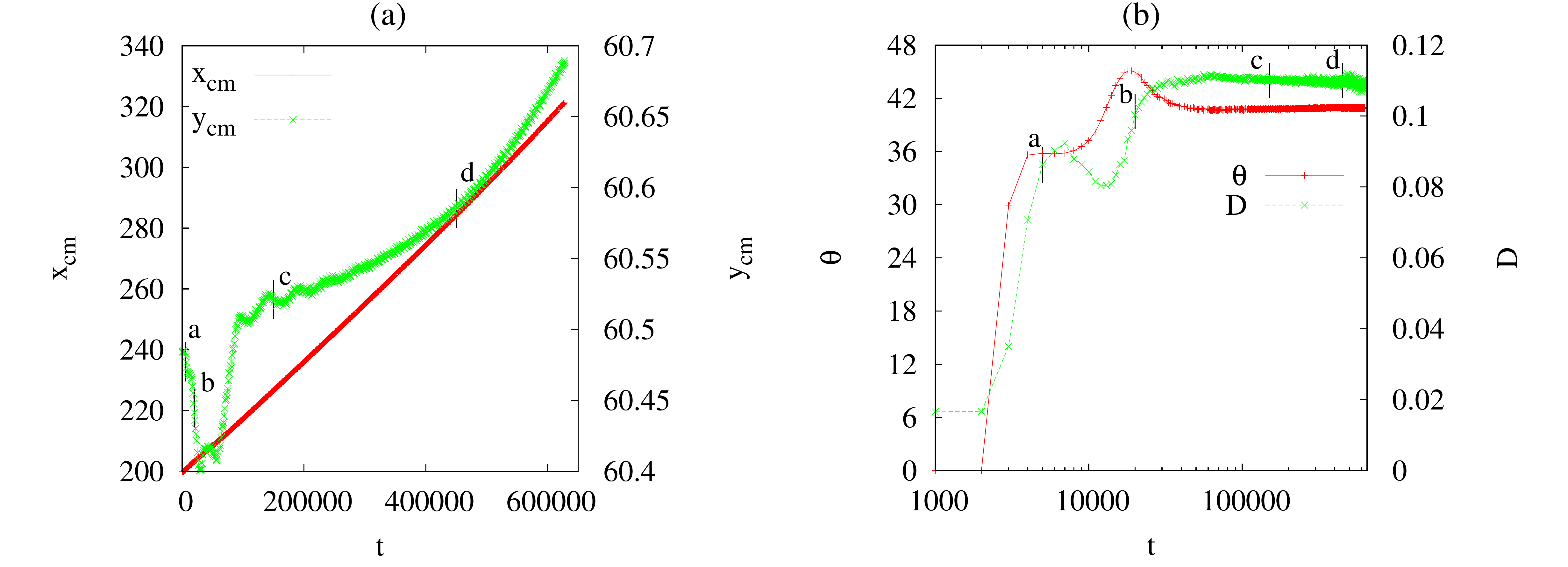}}
\caption{In panel (a) a plot of the $x$ and $y$ components of the centre of mass (red plusses-left scale and green crosses-right scale, respectively) of the droplet of Fig.~\ref{fig5} are reported. 
The droplet acquires motion 
with an almost constant speed since the shear is switched on.
In panel (b) the angle the major axis forms with the shear direction (red plusses, left scale) and the deformation parameter $D$ (green crosses, right scale)
are reported. Logarithmic time scale is set on the x-axis. 
The droplet attains a weakly deformed steady state with the major axis forming an angle of $\simeq 43$ degrees with the shear direction. The black lines in both panels
indicate the times at which snapshots (a)-(d) of Fig.~\ref{fig5} are taken.}
\label{fig6}
\end{figure*}

Due to the low value of the anchoring strength $L$, the dynamics discussed so far shows several similarities with the one observed in a single isotropic droplet (binary fluid-like) under shear.
On the other hand, when $\lambda\simeq 1$ the droplet attains a nutshell-like steady state in which two defects are located on its surface (see Fig.~\ref{fig1}b). This significatively affects its dynamics under shear, as shown in Fig.~\ref{fig5} and in  Movie S1.  If $\dot{\gamma}=1.67\times 10^{-4}$, the droplet, similarly to the previous case, stretches and elongates along the shear direction
(Fig.~\ref{fig5}a-b). 
The tilt of the major axis of the droplet can be measured by looking at the time evolution of the angle $\theta$ it forms with the shear direction, as reported in Fig.~\ref{fig6}b (left scale): the droplet orientation angle initially achieves a maximum at $\simeq 45$ degrees and afterwards relaxes to a constant value of $\simeq 41$ degrees.
Notice that the value of the angle at the steady state is very similar to the one measured for the $\lambda\ll 1$ case where defects are absent.
During their rotational relaxation the two defects migrate in phase along the surface, but out of phase with the rotational motion of the major axis of the droplet, before getting pinned on opposite sides at the steady state.
This difference might be ascribed to the flow generated by the symmetric shear. This stretches the droplet and changes the local orientation of the director field near the defects, pushing them along opposite directions. While the axis connecting them rotates of an angle larger than $90$ 
degrees around the centre of mass of the droplet, that of the droplet aligns along the flow direction, almost parallel to the director field in the bulk.
It is worth noting that defects are located on the surface of the droplet but not belonging to it as they are part of the liquid crystal phase. 
In addition the droplet acquires unidirectional motion along the $x$-axis with an almost constant speed (see Fig.~\ref{fig5}c-d) as the position of the $y$ component of the centre of mass
is slightly shifted upwards (see Fig.~\ref{fig6}a in which the $x$ and the $y$ components of the centre of mass are reported), and attains a final steady state whose shape is only weakly deformed by the shear (Fig.~\ref{fig6}b, green crosses, plots the deformation parameter $D$). We call it bound state (BS), as both defects remain on the droplet surface.

\begin{figure*}[htbp]
\centerline{\includegraphics[width=1.0\textwidth]{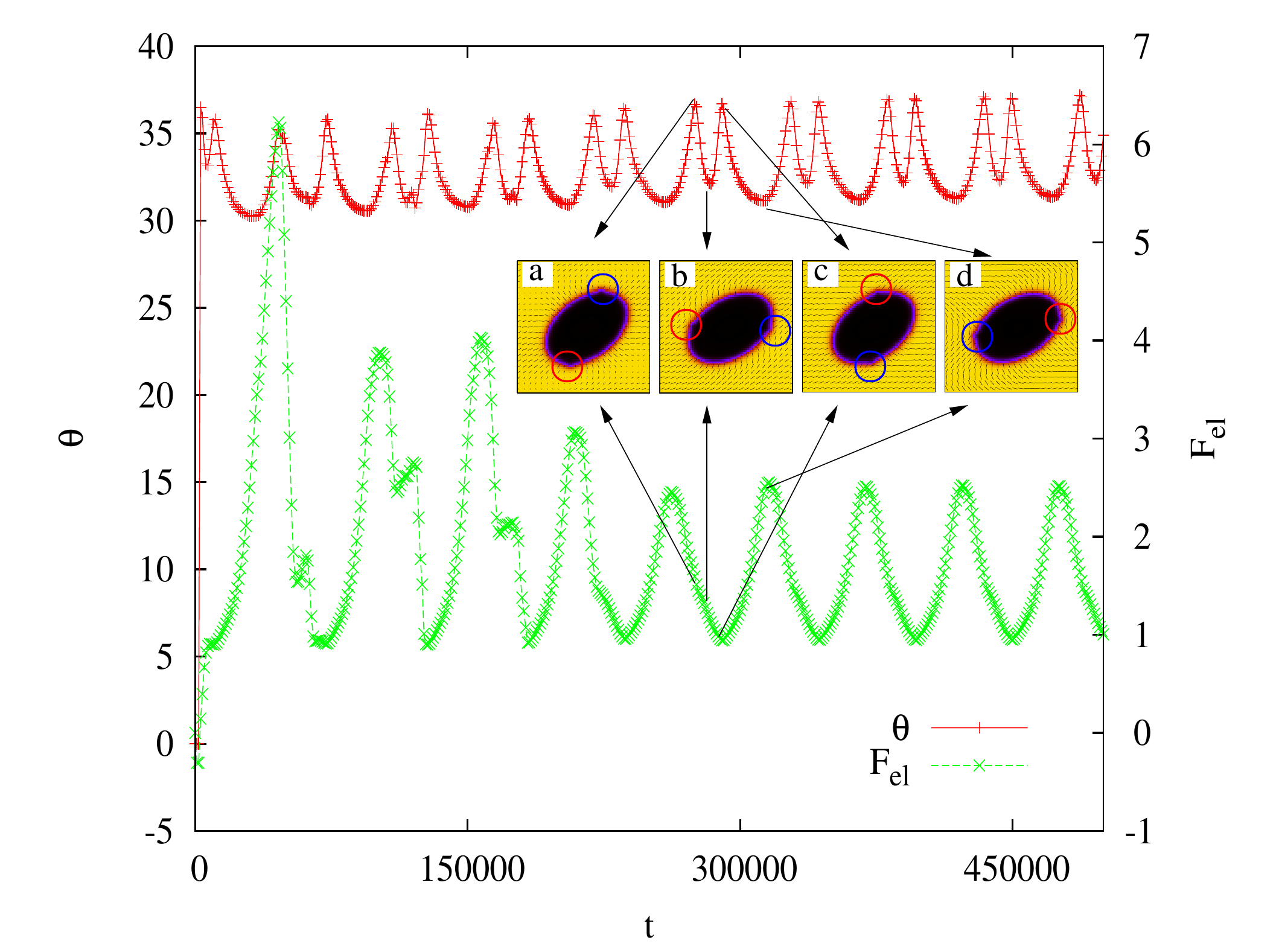}}
\caption{A plot of the angle $\theta$ the major axis of the droplet forms with the shear direction (red plusses, left scale) and a plot of the elastic liquid crystal free-energy
(green crosses, right scale) versus simulation time are reported for the dynamics shown in Movie S2. 
The angle $\theta$ displays a periodic behaviour in which two close maxima around $38$ degrees correspond to the dynamic state
of both defect close to the walls, whereas at the local minima defects are far from walls. The elastic free-energy (the sum of the elastic term in the bulk free energy multiplied by the constant $K$ 
and the surface anchoring contribution multiplied by the constant $L$) oscillates too. The four insets show the droplet and the nearby director profile. Red and blue circles 
indicate the position of the defects. Here $A_{iso}/A_{lc}\simeq 2.7\times 10^{-2}$ at steady state.}
\label{fig7}
\end{figure*}

A persistent rotation of the defects can be achieved if the shear rate is further increased. Movie S2 shows the dynamics with $\dot{\gamma}=4.2\times 10^{-4}$.
Similarly to the previous case, the droplet quickly deforms and aligns along the shear direction while both defects rotate clockwise. However, due to high value of $\dot{\gamma}$, the velocity field (very intense near the walls but weak in the centre, where the typical vortex pattern inside the droplet forms) pushes the defects over the position attained for lower $\dot{\gamma}$ and drives them around the entire surface of the droplet. 
In particular, during an entire cycle,  they speed up their motion near the walls of the cell and then, at the end of each cycle, slow down as their respective distance from the walls augments. 
This dynamics also leaves temporary wake-like signatures on the director field departing from the defects (red wakes in Movie S2) and short-living opposite charge topogical defects which annihilate quickly (red spots appearing in the nematic phase, see Movie S2). 
A measure of the angle $\theta$ that the major axis forms with the shear direction is reported in Fig.~\ref{fig7} with a plot of the elastic free-energy
\begin{equation}
F_{el} = \frac{K}{2}(\partial_{\alpha}Q_{\beta\gamma})^2+L(\partial_{\alpha}\phi)Q_{\alpha\beta}(\partial_{\beta}\phi).
\end{equation}
At long  times the angle $\theta$ displays two close local maxima, stabilized around $38$ degrees, spaced out by two local minima, one short and one large, both around $30$ degrees. 
Although defects continuously rotate on the surface of the droplet the major axis exhibits a characteristic oscillatory behaviour, in which the two minima are achieved when defects slow down their motion (far from the walls) and the two close maxima during the successive cycle (near the walls) (see Fig.~\ref{fig7}, red plusses, left scale). 
The elastic free energy of the nematic oscillates as well at long times  (see Fig.~\ref{fig7}, green crosses, right scale); in particular its local maximum corresponds roughly to the large minimum of $\theta$ and its minimum to the local maximum (on its left) of $\theta$ observed during a cycle. 
A crude explanation of this can be arguably related to the backflow: the velocity field, much higher near the walls than in the middle of the cell, aligns the director field more strongly when defects are closer to the walls (hence diminishing the elastic free energy) then far from them. Indeed, the director profile observed in these states supports this interpretation (see the insets (c) and (d) of Fig.~\ref{fig7}). On the other hand, the director field at the other extremes of the angle $\theta$ is significatively different (see the insets (a) and (b) of Fig.~\ref{fig7}). In particular the state (b) has a lower elastic free energy than that in (d), meaning that elastic deformations are weaker in (b) than in (d) (notice that these states correspond to the two minima of $\theta$) whereas the elastic free energy of the state (a) has a similar value of that in (b).
Interestingly, both the director and the corresponding velocity field (in the state (a)) acquire an out of plane component (along the $z$-direction) which accompanies the escape of the cores of the defect pair into the third dimension. 
The out of plane component of the velocity field in particular is roughly one order of magnitude lower than the other two components.
Unlike a droplet-free flow aligning nematic liquid crystal, the sole presence of topological defects on the droplet surface unveils an unexpected flow-tumbling-like dynamics which is overall akin to that observed in a strictly two-dimensional system but for a slightly higher value of the shear rate. In the latter in particular both the director and the velocity field are not significatively different from  those observed in the quasi-2d case, and the final steady state (the oscillatory bound state) is preserved.
In addition, at the steady state the rate of deformation $D$ of the droplet weakly oscillates around $0.2$ and the capillary number is $\simeq 0.28$.
The steady state described above, in which the shear stress induces an oscillatory-rotational motion of the antipodal defects pair close to the droplet, is, to our knowledge, a new result for an inverted nematic  emulsion and we call it the {\sl oscillatory bound state} (OBS).

\begin{figure*}[htbp]
\centerline{\includegraphics[width=0.92\textwidth]{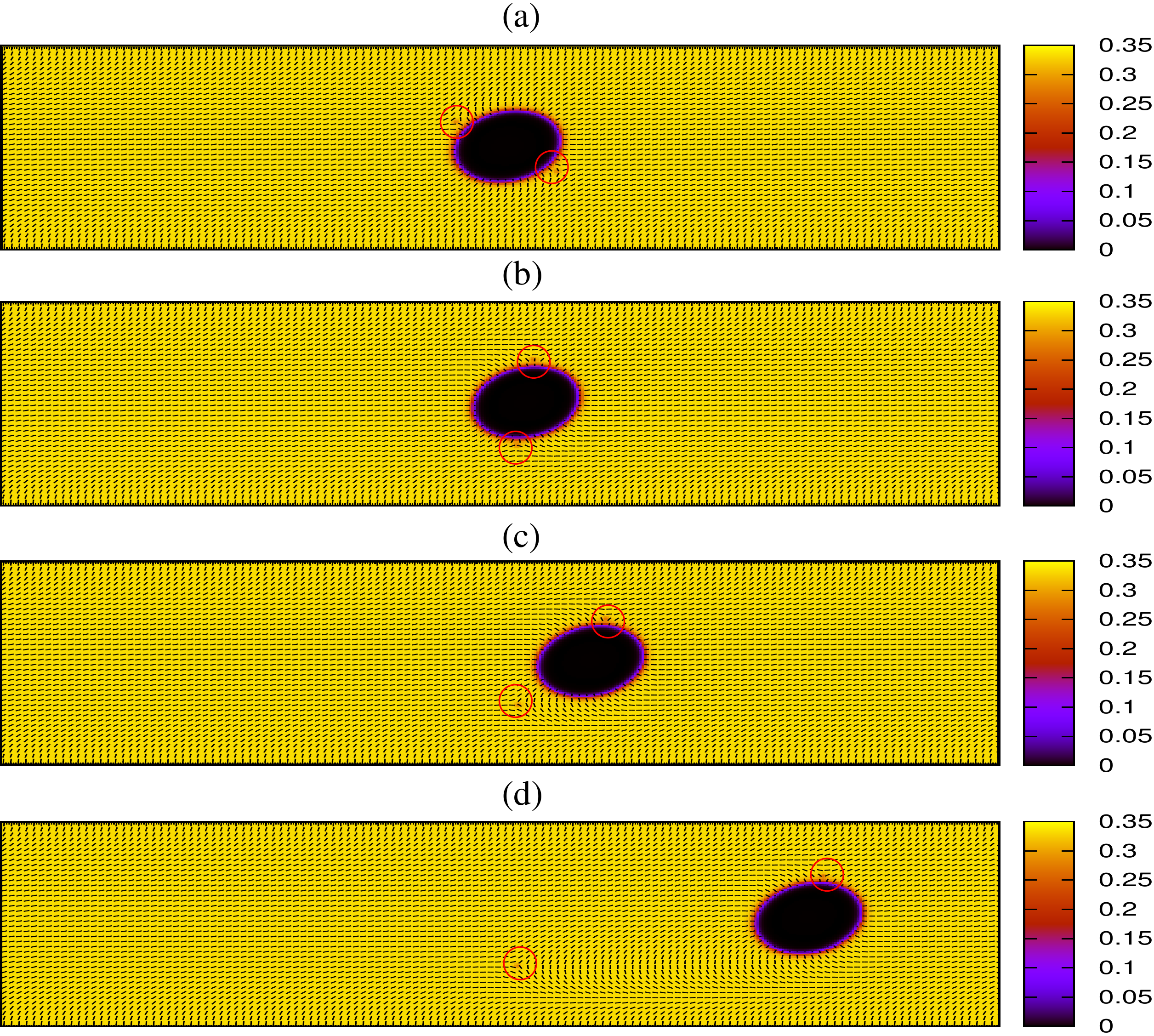}}
\caption{Dynamic evolution of a droplet of radius $R=22$ with defects located out of its surface with $\dot{\gamma}=1.25\times 10^{-4}$ (see Fig.~\ref{fig1}c or d for the corresponding equilibrium state). The droplet is stretched along the shear flow and the two defects, initially on opposite side of the droplet (a), move in clockwise direction along its surface (b). Afterwards the one near the bottom wall detaches from the surface (c) and migrate away from it (d),  leaving a characteristic comet-like signature on the director field, whose bulk structure is now almost completely aligned along the shear flow. At steady state the capillary number is $\simeq 0.095$ and $A_{iso}/A_{lc}\simeq 3.4\times 10^{-2}$. See also Movie S3 for the complete dynamics. }
\label{fig8}
\end{figure*}

\begin{figure*}[htbp]
\centerline{\includegraphics[width=1.05\textwidth]{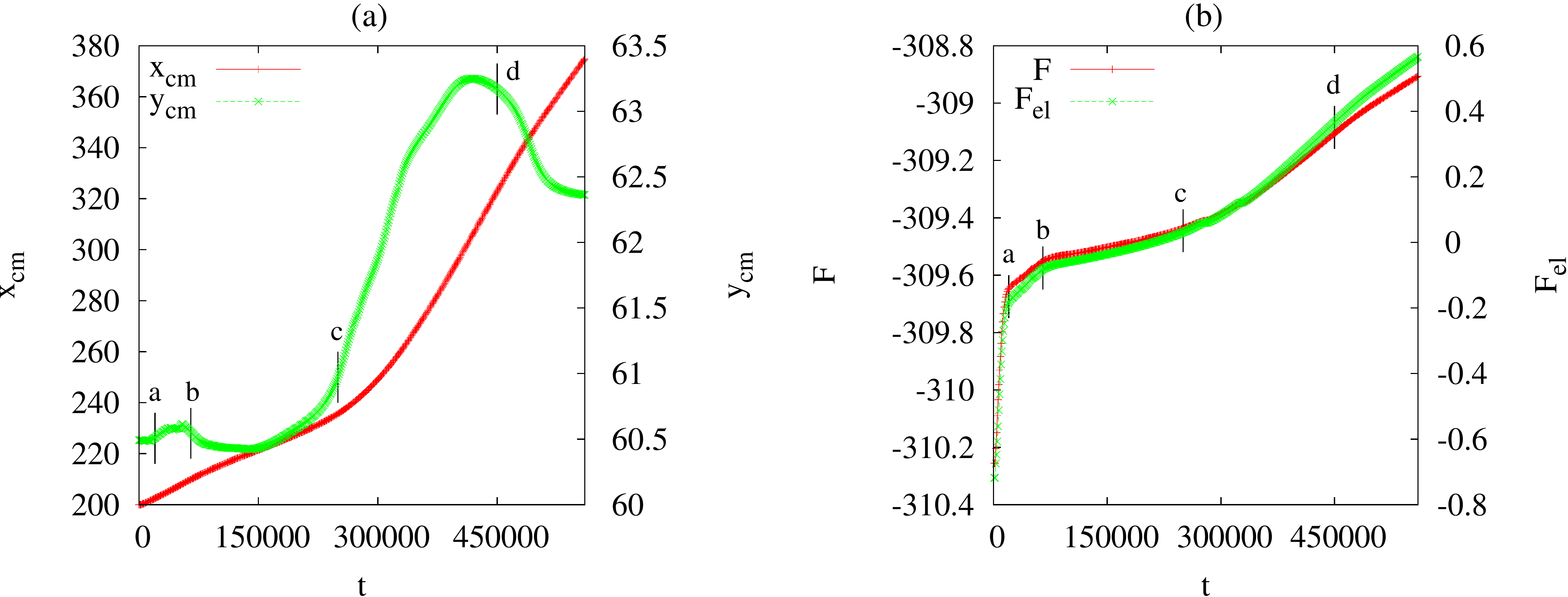}}
\caption{(a) $x$ and $y$ components of the centre of mass (red plusses, left scale and green crosses, right scale respectively) relative to the dynamics shown in Fig.~\ref{fig8}. 
After the shear is switched on the droplet acquires unidirectional motion along the $x$-axis, with an almost constant velocity at early times, up to $t\simeq 250\times 10^{3}$. Afterwards, when the lower defect detaches from the droplet, this moves faster attaining
a second steady state with a higher velocity. On the other hand the $y$-component remains almost constant although initially, when the droplet is at rest, slightly shifted upwards. This breaks the symmetry and allows the droplet to move along a preferential direction. (b) A plot of the total free energy (red plusses, left scale) and the elastic one (green crosses, right scale) 
for the dynamics shown in Fig.~\ref{fig8} is reported. Black lines indicate the times at which the corresponding snapshots of Fig.~\ref{fig8} are taken.} 
\label{fig9}
\end{figure*}

When $\lambda \gg 1$ the defects of the equilibrium configuration are further apart from the droplet surface (see Fig.~\ref{fig1}c-d): this couples to the imposed shear to give rise to further novel dynamical behaviours.
In particular, depending on the value of the shear rate, one can identify three additional dynamical regimes: for very low shear rates (tipically $\dot{\gamma}<10^{-5}$) both defects remain close to the droplet surface, as in the oscillatory bound state; for intermediate shear rate only one defect goes away from the droplet ({\sl single bound} (SB) state)  and finally, for sufficiently high shear rates,  both defects move into the bulk~\footnote{For extremely high shear rates the isotropic phase disappears as usual.} ({\sl unbound} (U) state).
In Fig.~\ref{fig8} and in Movie S3 we report the dynamics of the SB state.
For $\dot{\gamma}=1.25\times 10^{-4}$ the droplet initially behaves similarly to what seen for the BS case where
both defects, located on opposite sides along the equator, rotate simultaneously in the clockwise direction (Fig.~\ref{fig8}a-b). Later on the rotation is arrested and the droplet-defect system, dragged by the fluid flow, slowly moves unidirectionally along the $x$-axis (Fig.~\ref{fig8}c), whereas the defect near the bottom wall gradually leaves the droplet and moves into the bulk along the opposite direction (from right to left in the bottom half part of the system)  (Fig.~\ref{fig8}d). 
This defect motion leaves a characteristic comet-like signature of the director field  in the bulk. Note that the SB state of Fig.~\ref{fig8}d occurs because, at equilibrium (i.e. before the shear is switched on), the droplet centre of mass is not equidistant from the two walls ($y_{cm}\simeq 60.5 > L/2$ as shown in Fig.~\ref{fig9}a). This is the reason why, under shear, the droplet is slowly dragged by the flow along the $x$-direction and, more importantly, reduces the distance between its surface and the top defect, increasing at the same time the distance from the bottom defect, which is then pushed backwards by the fluid.
It is interesting to note that, after the bottom defect moves into the bulk, the droplet accelerates attaining  a novel steady state with a higher velocity.
During this process the total and the elastic free energies show a very similar behaviour: at early times (when defects just rotate around the droplet) they increase rapidly. At intermediate times instead they increase very slowly  (this regime refers to the situation in which the defects move slowly along opposite direction parallel to the walls). Finally, at longer times,  when the droplet and the bottom defect move apart, they increase quite rapidly again (see Fig.~\ref{fig9}b). 

\begin{figure*}[htbp]
\centerline{\includegraphics[width=0.92\textwidth]{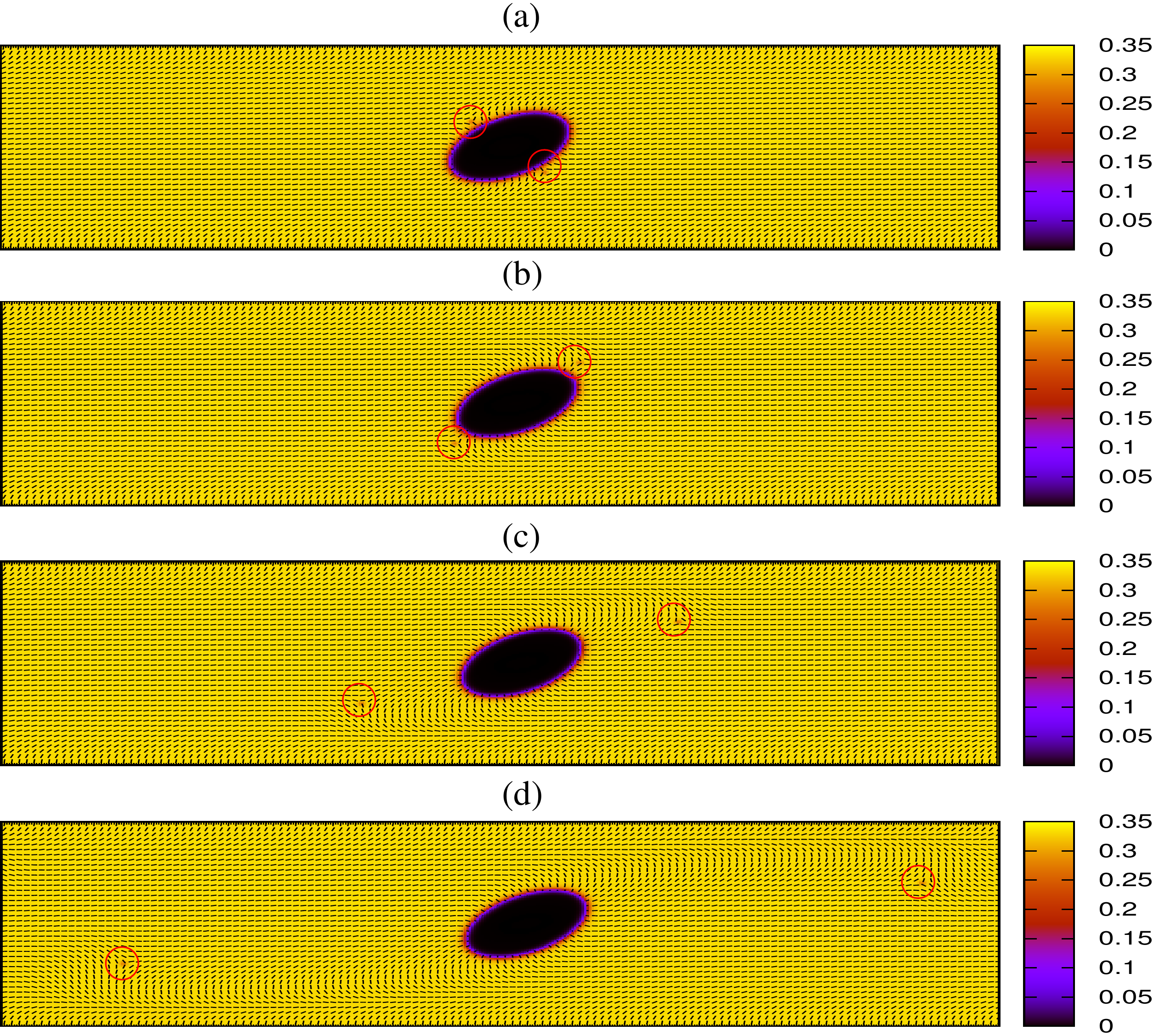}}
\caption{Dynamic evolution of a droplet of radius $R=22$ with defects located out of its surface with $\dot{\gamma}=4.2\times 10^{-4}$
(see Fig.~\ref{fig1}c or d for the corresponding equilibrium state). 
The dynamics at early times ((a) and (b)) is analogous to that of the SB state, except for a larger droplet deformation $D=0.2$. Indeed, the droplet stretches along the
shear flow and defects simultaneously rotate clockwise. Afterwards, due to the high shear rate, both defects are dragged away from the
surface of the droplet, whose position, unlike the SB case, is not appreciably changed. At steady state the capillary number is $\simeq 0.28$ and $A_{iso}/A_{lc}\simeq 2.6\times 10^{-2}$.}
\label{fig10}
\end{figure*}

A state in which both defects migrate into the bulk (U, unbound, state) is achieved by further  increasing the shear rate.  In Fig.~\ref{fig10} and in Movie S4 we show this situation ($\dot{\gamma}=4.2\times 10^{-4}$). The early times dynamics is similar to that observed for the  SB state (see Fig.~\ref{fig10}a-b), except for a larger deformation of the droplet ($D\simeq 0.2$ here while  $D\simeq 0.08$ in the SB state of Fig.~\ref{fig8}). However, the shear rate is now intense enough along the $x$-direction to detach both defects from the droplet. 
Interestingly, there is not an appreciable motion of the droplet along the shear direction until the defects, reappearing at the periodic side of the lattice,  reapproach the droplet again. It is worth noticing that an unbound state could be also achieved by suitably controlling the elasticity of the liquid crystal. Indeed, we have found a very similar dynamics for a droplet with a smaller radius ($R=15$) with the parameters of Fig.~\ref{fig1}d (hence with $K=8\times 10^{-3}$) but with $\dot{\gamma}=1.25\times 10^{-4}$.

\subsection{Diagram of non equilibrium steady states}

The results just presented  can be summarised and rationalised into a  phase diagram in which the observed steady states are reported in the $\lambda-\dot{\gamma}$ plane.
The diagram displays different dynamical responses of the droplet under a linear shear.
For $\lambda\ll 1$, when defects are not present,  a dynamical response typical of a Netwonian emulsion  dominates 
for all values of shear rates $\dot{\gamma}$.  In particular if $\dot{\gamma}\le 10^{-3}$ the droplet deforms and elongates
along the shear flow whereas, for higher values of $\dot{\gamma}$, it breaks into smaller droplets and 
eventually evaporates\footnote{We borrow this term from the liquid-gas thermodynamics as the shrinkage
of the isotropic phase is due to the shift of the temperature of the 
isotropic-nematic transition which favours the liquid-crystal phase.}. 
If, on the other hand, $\lambda\simeq 1$, defects pair forms very close to the droplet surface 
and, due to  the interplay between their dynamics and the deformation-rotation dynamics of droplet, 
three new interesting steady states can be identified.
For small shear rates defects are bound in proximity of the surface of the droplet and no appreciable
difference with the standard binary fluid behaviour is observed.  Increasing $\dot{\gamma}$ unveils a region in which the
motion of the droplet  along the shear direction combines with a partial rotation of the defects along its surface.
If $\dot{\gamma}$ is further increased, a second type of steady state occurs, in which defects are still bound to the droplet 
surface but now persistently rotate around it. Clearly, for sufficiently high values of $\dot{\gamma}$, the droplet breaks up 
and eventually evaporates.
For $\lambda\gg 1$ the defects pair forms still in proximity of the droplet but in the nematic phase 
and this gives rise to two additional steady  states: in the first one, for intermediate values
of $\dot{\gamma}$,
only one defect remains close  to the droplet while the other one starts to move freely with the shear flow 
(single bound state). When, on the other hand, $\dot{\gamma}$ is sufficiently high, the second defect
leaves the droplet as well and moves in the bulk (unbound state).

\begin{figure*}[htbp]
\centerline{\includegraphics[width=1.0\textwidth]{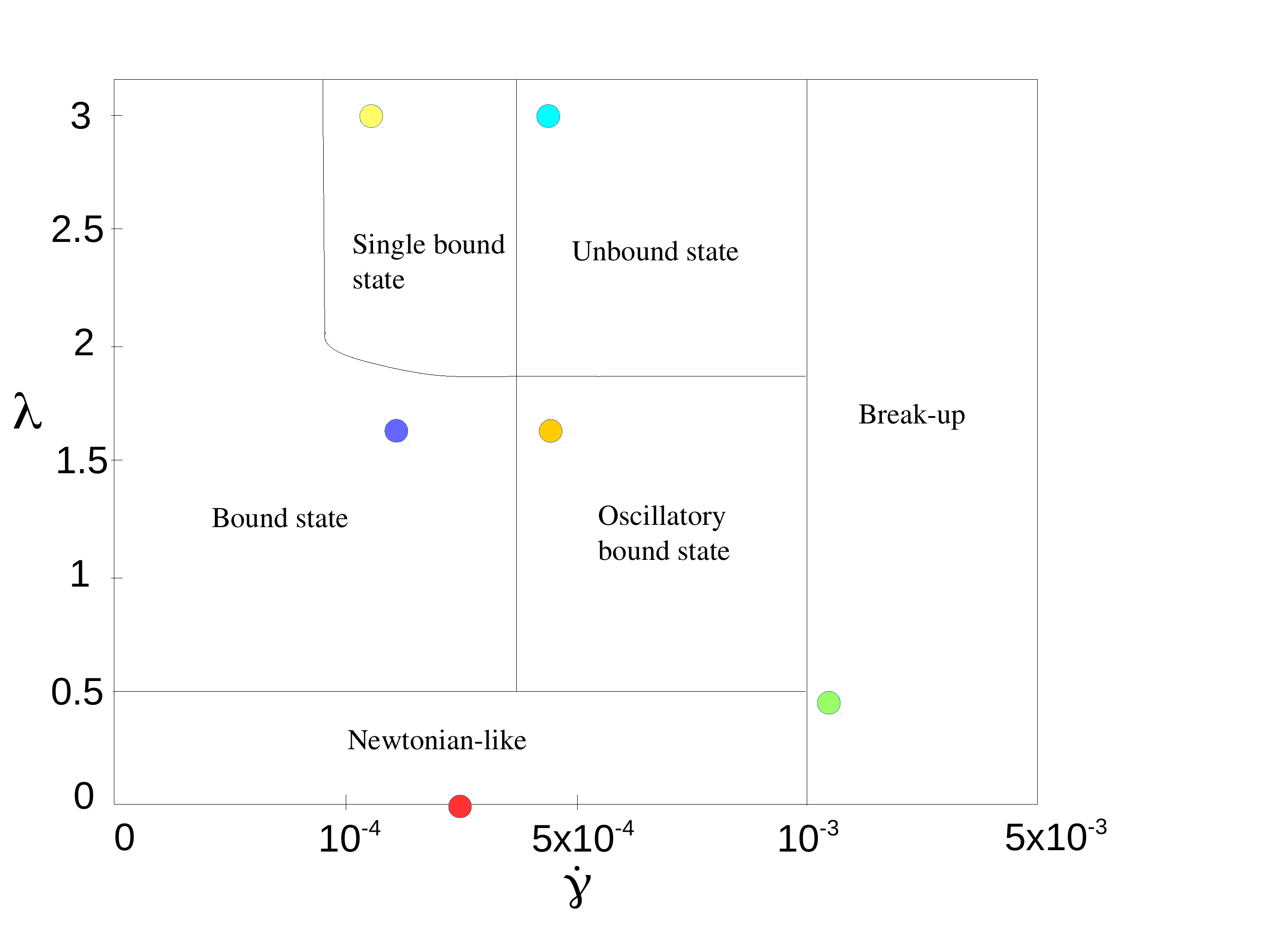}}
\caption{A qualitative phase diagram of the steady states observed under shear in the $\lambda-\dot{\gamma}$ plane (both in simulation units).
For $\lambda \ll 1$ defects are absent and a Newtonian dynamics dominates. For $\lambda \simeq 1$ defects pair
forms near the surface of the droplet. For small and intermediate values of $\dot\gamma$ the droplet is either quiescent or acquires motion
along the shear direction combined with a partial rotation of the defects. This is the bound state (BS). If $\dot\gamma$ is further
augmented, a persistent rotation of both defects is found. This is the oscillatory bound state (OBS). 
For $\lambda\gg 1$ defects are fully in the nematic phase. For intermediate values of $\dot\gamma$ one of the defects
leaves the droplet while the other one remains connected (single bound state (SB)). For higher values of $\dot\gamma$
both defects are dragged by the shear flow and disconnect from the droplet (unbound state (US)).  In all cases, for very
high $\dot\gamma$, the droplet breaks into smaller droplets which eventually evaporate.
The circles indicate the position in the phase diagram of the systems described in the figures of the manuscript.
More specifically we have: $\lambda=0$,  $\dot{\gamma}=2.5\times 10^{-4}$ (red circle); $\lambda\simeq 0.49$, $\dot{\gamma}=1.25\times 10^{-3}$(green circle);
$\lambda\simeq 1.58$, $\dot{\gamma}=1.67\times 10^{-4}$ (blue circle); $\lambda\simeq 1.58$, $\dot{\gamma}=4.2\times 10^{-4}$ (orange circle); 
$\lambda\simeq 3$, $\dot{\gamma}=1.25\times 10^{-4}$ (yellow circle); $\lambda\simeq 3$, $\dot{\gamma}=4.2\times 10^{-4}$ (cyan circle).}
\label{fig11}
\end{figure*}

\subsection{Homogeneous anchoring}
When homogeneous (or tangential) anchoring is set on the surface of the droplet a different dynamical response of the system is observed.
For these cases, our simulations are performed by using a rectangular box of size $Lx=400$, $Ly=80$ with a droplet of radius $R=15$.
When defects are imaginary (see Fig.~\ref{fig2}a), the dynamical behaviour under shear is similar to the one observed with homeotropic anchoring at low shear rate ($\dot{\gamma}=2.5\times 10^{-4}$): the droplet aligns with shear flow  with  $\theta\simeq 40$.
In the presence of defects at the droplet surface  (as in Fig.~\ref{fig2}b), however, the situation is different from that with homeotropic anchoring. In this case, we observe a slow drift of the droplet along the $x$-direction induced by the defect at its right. Accordingly, the droplet  increases its distance from the defect located on its left, which gradually leaves the surface of the droplet (see Movie S5). This establishes a transient single-bound state which lasts until the ``free'' defect wraps the periodic boundaries: when this happens, the mobile defect rotates clockwise around the surface and this time remains bound, trapped by the elastic interactions. A similar dynamics has also been observed for higher shear rates, before the droplet evaporates. Steady unbound states with defects in the bulk of the nematic host could instead be found either by decreasing the nematic elastic constant $K$ or by increasing the surface anchoring $L$. Indeed, when imposing $L$$=5.6\times 10^{-2}$ the equilibrated droplet looks very similar to the one of Fig.~\ref{fig2}; however, under a shear rate of $\dot{\gamma}=2.5\times 10^{-4}$ we now observe defect detachment. This is possible because a large value of $L$ augments the distance between each defect, which favours detachment. 

\subsection{Extension to 3D systems}
The rich phenomenology found in the 2D system may be observed in thin film of inverted nematic emulsions under shear; however it  is of interest to ask to which extent this also occurs in a fully 3D system. For simplicity, we only consider here the case where without shear the droplet is accompanied by a Saturn ring defect (the case of a hyperbolic hedgehog defect may also be of interest).  
Previous experimental and numerical studies on micron-size droplets hosted in a liquid crytstal and advected by a flow indicate that the Saturn ring defects formed around them are visibly displaced in the downstream direction and eventually collapse into a hyperbolic point defect~\cite{Zhou_et_al_JFM_2007,Khullar_et_al_PRL_2007}.
Here we look at how a linear shear may impact on the Saturn ring-droplet system by restricting ourselves to 3D isotropic droplets hosted in a liquid crystal fluid that is homeotropically anchored to the droplet surface. The system is made by a single droplet of radius $R=14$ sandiwched between two walls at $L_z=0$ and $L_z = 40$.  Along the $x$ and $y$ directions, periodic boundary conditions are considered  with $L_x=60$ and $L_y=40$. Since, as a first approximation,  the droplet is expected to assume a generic ellipsoidal shape, we compute a set of three dimensionless numbers, $c_l$, $c_s$ and $c_p$, measuring respectively the degree of prolateness, sphericity  and  oblateness of the ellipsoid.

These quantities are defined as 
\begin{eqnarray}
c_l&=&\frac{a-b}{a+b+c},\\
c_s&=&\frac{3c}{a+b+c},\\
c_p&=&\frac{2(b-c)}{a+b+c},
\end{eqnarray}
where $a$, $b$ and $c$ are the principal axes of the ellipsoid, such that $a\ge b\ge c\ge 0$. 
For example, to establish whether the ellipsoid is more prolate than oblate one can compare $c_l$ with $c_p$: if $c_l > c_p$, the ellipsoid tends to be prolate whereas if $c_l< c_p$ the droplet is more oblate. 
Two extreme cases can be identified: for $c_l=1$ the ellipsoid degenerates into a segment while, if $c_p=1$ (i.e. $a=b > 0$ and $c=0$) the ellipsoid degenerates into a two-dimensional circle. Finally if $c_s = 1$ the spherical shape is recovered  ($a=b=c$).

Since, similarly to the 2D case, we expect that the equilibrium location of the defect and its dynamics under shear flow depends on the adimensional ratio $\lambda$, we consider the two extreme cases of $\lambda\simeq 1$ and $\lambda >> 1$.
In both cases the system is first let to equilibrate in absence of shear; the resulting configuration is then used as initial condition of the shear experiment.

\begin{figure*}[htbp]
\centerline{\includegraphics[width=1.15\textwidth]{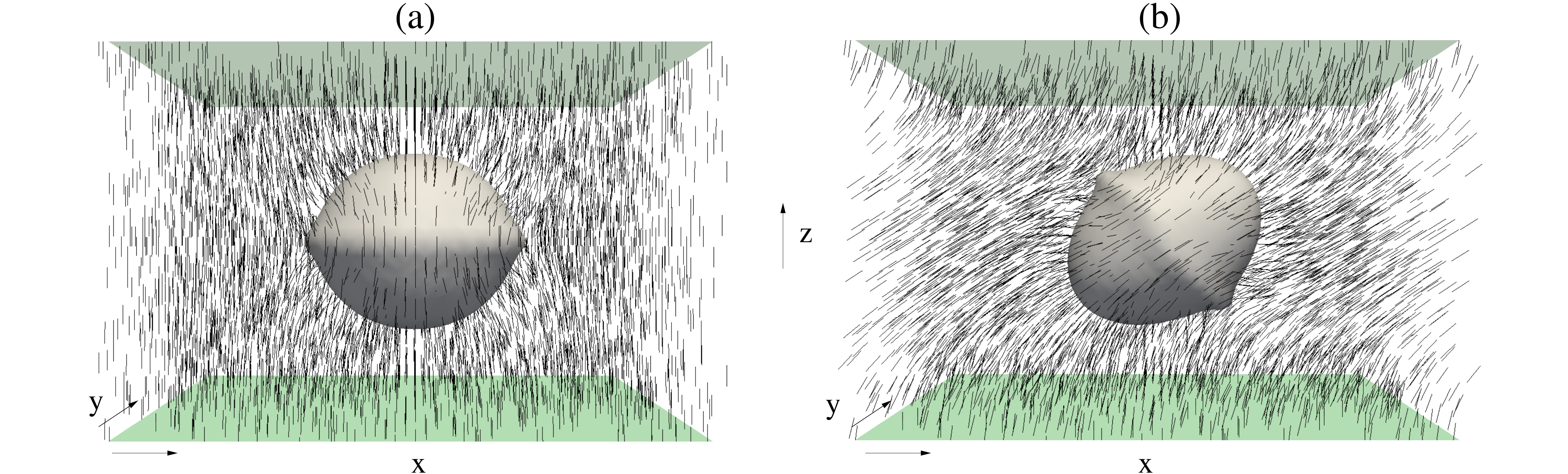}}
\caption{(a) Equilibrium configuration of an isotropic 3D droplet embedded in a nematic liquid crystal homeotropically anchored both at  the droplet surface and at the walls. The equilibrium value of the parameter $\lambda$ is roughly equal to $1$.
(b) Steady state of the system described in (a)  under a moderate  shear ($\dot{\gamma}=3.75\times 10^{-4}$). The droplet (namely its major axis) orients parallel to the shear flow while the Saturn ring (almost undeformed) rotates in the $y-z$ plane following the major axis of the droplet.
The director field is almost everywhere aligned along the shear direction except in proximity of both the droplet surface and  the walls, where strong hometropic anchoring is set. Here $K=5\times 10^{-2}$ and $L=-4\times 10^{-2}$.}
\label{fig12}
\end{figure*}

The equilibrium state for  $\lambda\simeq 1$ is shown in Fig.~\ref{fig12}a: it is readily seen that the droplet  assumes a slightly deformed spherical shape with $c_l\simeq 0.036$, $c_s\simeq 0.89$ and $c_p\simeq 0.073$. The nutshell-like shape of the droplet is due to the location of the Saturn ring defect right at its surface, a situation similar to that reported in Fig.~\ref{fig1}b  for the corresponding 2D case.
Note that the director field is almost everywhere parallel to the $z$-axis, except in proximity of the  droplet surface where, due to the strong homeotropic anchoring, a large splay-bend distortion emerges (Fig.~\ref{fig12}a). 
When a moderate shear flow is imposed on the system ($\dot{\gamma}=3.75\times 10^{-4}$),
the droplet does not move with the flow but simply stretches and elongates along the shear flow (resembling the 2D case of Fig.~\ref{fig5}), and achieves a final steady state whose shape is that of a prolate ellipsoid ($c_l\simeq 0.12$), with its  major axis oriented at roughly $33$ degrees with the shear direction (Fig.~\ref{fig12}b  and Movie S6).

As expected, the director field orients preferentially along the shear direction at the centre of the system while it strongly bends both in proximity of the 
Saturn ring  and towards the walls.
The Saturn ring itself does not experience an appreciable deformation but rotates in the $y-z$ plane with its symmetry axis (passing through the centre of the ring) remaining almost parallel to the major axis of the ellipsoid.

\begin{figure*}[htbp]
\centerline{\includegraphics[width=1.1\textwidth]{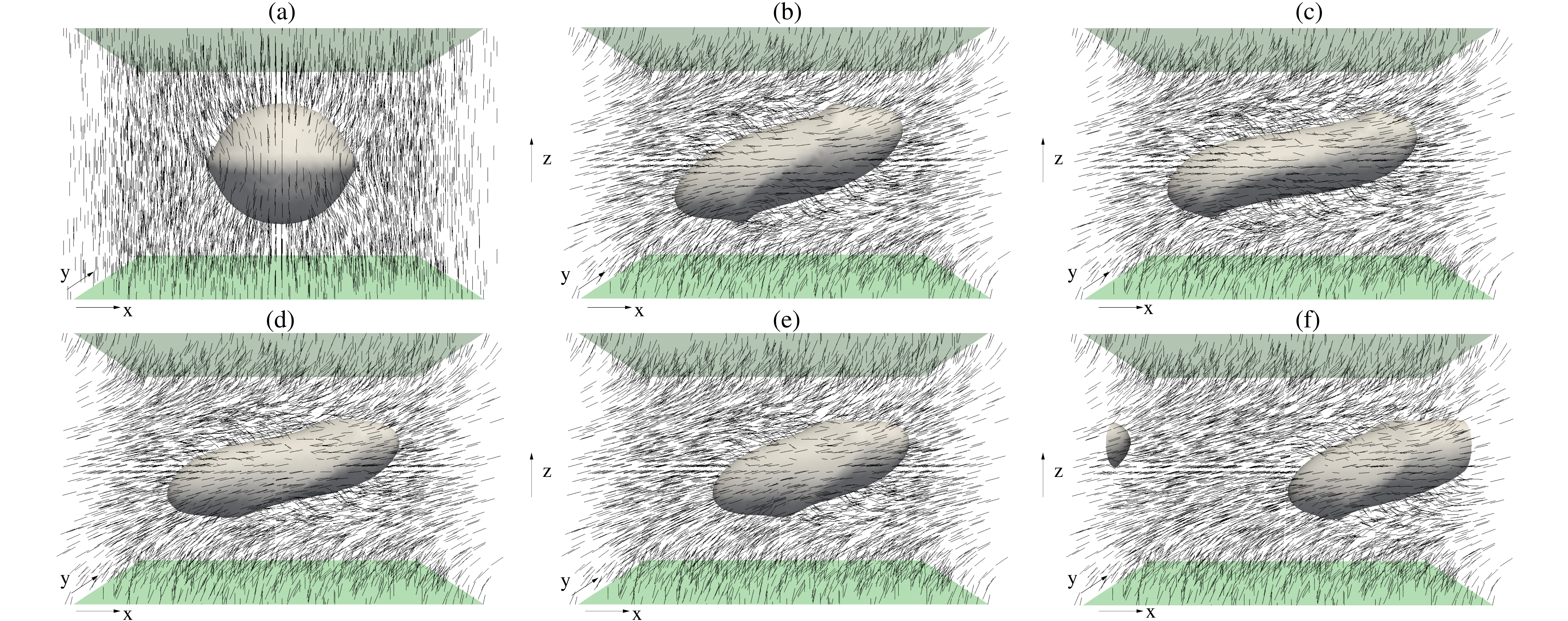}}
\caption{(a) Equilibrium configuration of a 3D isotropic droplet embedded in a nematic liquid crystal homeotropically anchored both at the droplet surface and at the confining walls. 
At equilibrium $\lambda\simeq 1$. 
Panels (b-d) represent intermediate states of the system when it is subject to a sufficiently strong shear rate ($\dot{\gamma}=1.1\times 10^{-3}$).
When the shear is turned on, the droplet is initially stretched by the fluid (b) 
and afterwards it aligns along the direction of the shear flow (c). 
Later on it slightly rotates around its major axis and moves upwards (along the $z$-axis), towards regions of higher shear flow (d). 
The droplet then acquires unidirectional motion along the $x$-axis and the Saturn ring moves downstream, i.e., in the opposite direction of the  motion (e-f). Here  $K=5\times 10^{-2}$ and $L=-4\times 10^{-2}$.}
\label{fig13}
\end{figure*}

For higher values of the shear rate the dynamics of the droplet and accompanying Saturn ring is significatively different. In Fig.~\ref{fig13} we report a simulation in which $\dot{\gamma}=1.1\times 10^{-3}$ (see also Movie S7). 
After the shear is switched on, the droplet initially aligns and elongates along the flow direction (Fig.~\ref{fig13}b) achieving a highly prolate ($c_l\sim 0.18$) intermediate (non-steady) state (Fig.~\ref{fig13}c). In this state the Saturn ring remains firmly anchored at the droplet surface. 
Interestingly, though, instead of surrounding the equator of the droplet (as for the case when $\dot{\gamma}=3.75\times 10^{-4}$), the disclination ring follows the droplet deformation and stretches along the entire surface.
Later on a more complex rearrangement is observed: the droplet moves slightly upwards (along the $z$-axis), in regions of the system in which the flow (direct along the $x$-axis) is more intense, and rotates around its major axis (Fig.~\ref{fig13}d). 
Finally it is pushed forwards and the Saturn ring slips downstream, opposite to the direction of motion (Fig.~\ref{fig13}e-f). The downstream motion of the Saturn ring agrees with the previous studies on similar systems~\cite{Zhou_et_al_JFM_2007,Khullar_et_al_PRL_2007} and can be considered as the 3D concounterpart of the 2D bound state observed before (see Fig.~\ref{fig5}).

\begin{figure*}[htbp] 
\centerline{\includegraphics[width=1.1\textwidth]{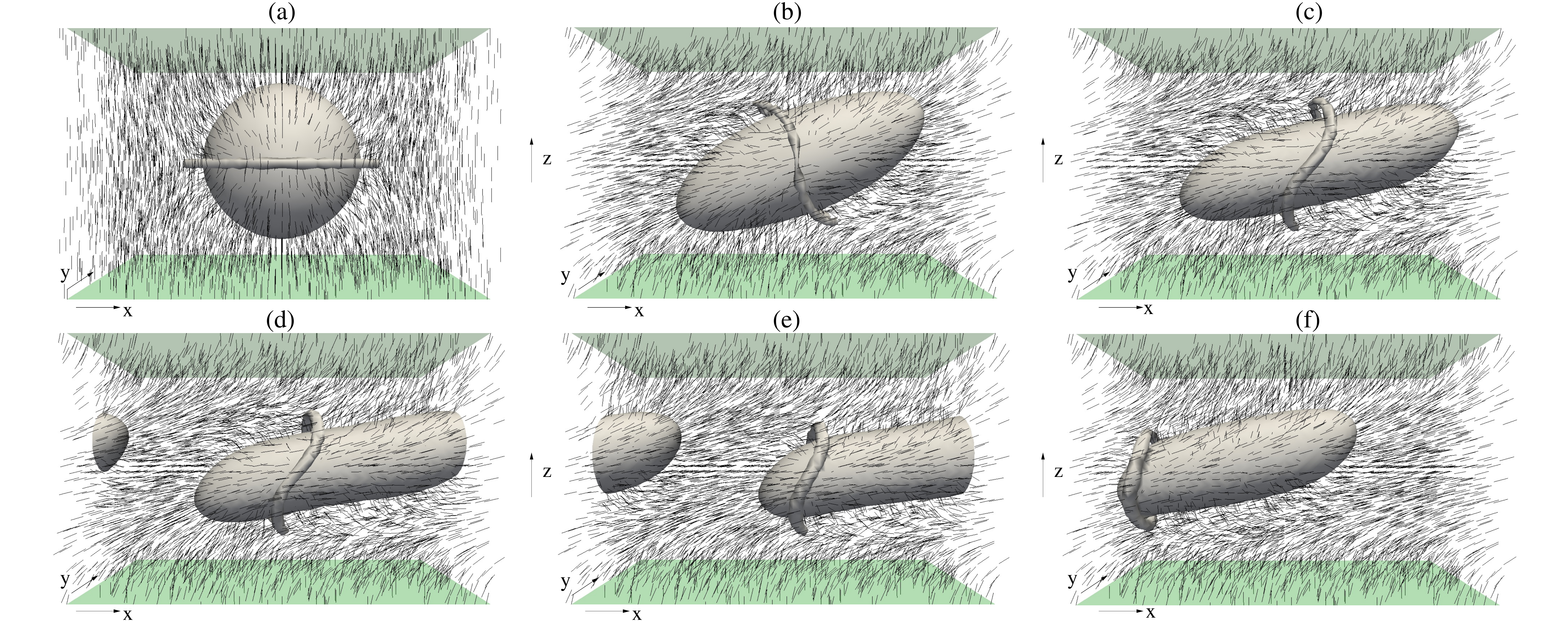}}
\caption{(a) Equilibrium configuration of a 3D isotropic droplet  embedded in a nematic liquid crystal homeotropically anchored both at the surface droplet and at the walls. Here $\lambda\gg 1$. 
When a sufficiently high value of shear is established ($\dot{\gamma}=10^{-3}$), 
the droplet aligns along the shear flow and elongates into an elliptic  prolate shape.
The Saturn ring, initially surrounding the equator of the droplet, rotates (with the axis almost parallel to the major axis of the emulsion) and slightly deforms ((b)-(c)). Later on, when the droplet is advected by the fluid along the $x$-axis, the Saturng ring moves downstream, and eventually gets pinned at the downstream extremity of the droplet ((d)-(e)-(f)).
Here  $K=5\times 10^{-2}$, $L=-8\times 10^{-2}$.}
\label{fig14}
\end{figure*}

For $\lambda\gg 1$ the equilibrium configuration is characterised by a Saturn ring surrounding the droplet but now fully embedded in the nematic phase (see Fig.~\ref{fig14}a).
As for $\lambda \simeq 1$ we can distinguish two dynamic regimes, depending on the value of the shear rate. While for a moderate shear rate the Saturn ring mantains its shape unaltered and simply orients its axis along the major axis of the deformed droplet,  for higher values a more complex behaviour is observed (see 
Fig.~\ref{fig14} and  Movie S8, in which $\dot{\gamma}=10^{-3}$).
Initially the droplet strongly deforms ($c_l\simeq 0.3$, $c_s\simeq 0.6$ and $c_p\simeq 0.08$) with its major axis aligning almost parallel to the shear flow,  while the Saturn ring, although located almost at the centre of the droplet, undergoes an $S$-like deformation (Fig.~\ref{fig14}b-c), less pronounced though than that observed for $\lambda\simeq 1$ and $\dot{\gamma}=1.1\times 10^{-3}$.
Afterwards, the droplet is advected by the flow along the $x$-axis while the disclination ring slips towards the rear part of the emulsion where eventually it gets pinned (Fig.~\ref{fig14}d-e-f).  Interestingly the shift of the Saturn ring occurs whenever the droplet acquires motion, namely when, for instance, its centre of mass shifts (either upwards or downwards).
Thus, unlike in 2d, where for $\lambda\gg 1$ by increasing the shear rate, the SB and the U state could be observed, here the defect ring sticks to the droplet even for intense shear rates. The further dimension therefore appears to diminish the number of possible dynamical states the system can explore.

Note that, although the starting position of the centre of mass of the droplet is placed at the centre of the simulation box, especially for high values of the shear rate this situation is highly unstable, and any small variations either in the director field or in the velocity field inevitably drags the droplet away from it.
For higher values of the shear rate we may expect the  Saturn ring to further shrink into a hyperbolic hedgehog, although we have not found evidence of this effect within the simulation times considered here. 

\section{Conclusions}

In conclusion, we have studied the dynamics of a 2D isotropic droplet immersed in a nematic host under an imposed symmetric shear flow.
The physics of the steady states is mostly determined by the shear rate and the strength of the anchoring of the director field at the droplet surface. 
For weak or no anchoring, the system behaves similarly to a droplet in a binary mixture of isotropic, Newtonian, fluids. On the other hand, for intermediate or strong anchoring a variety of non-trivial non-equilibrium steady states are possible. 
One option is to create an oscillatory steady state, where the droplet, for instance, tumbles in the flow. 
Another option is to create highly dynamic states where one or both the defects, which accompany the droplet when quiescent, detach from the surface of the droplet and become mobile due to the imposed shear.
For 3D systems and homeotropic anchoring the defect pair becomes a Saturn ring surrounding the droplet. We have shown that, regardless of the shear rate and of the surface anchoring strength, the disclination ring always remains localized around the droplet although it can be highly deformed and shifted downstream (similarly to the moving bound state observed in 2D). 
In this respect the unbound and single defect states observed in 2D can be seen as a limiting case of an inverted emulsion confined within a very thin film  of liquid crystal.

All these findings suggest that the rheological response of an isotropic droplet in a nematic host is much richer than previously probed, and can further be tuned by varying strength and nature of the anchoring, together with the magnitude of the imposed flow. 
A viable route to test our results experimentally is, for instance, in microfluidic and rheology experiments on water-liquid crystal emulsions (or other mixtures contaning a liquid crystal and a Newtonian fluid).

\section{Appendix 1}

Here we briefly describe how we calculate the perimeter of the droplet when its shape is deformed under shear. More specifically,
our simulations show that the droplet aligns along the direction of the shear flow, elongates and often acquires motion, with a shape
reminding that of an ellipse. In order to calculate the ellipse which best reproduces the boundary of the droplet, we compute the inertia tensor of 
a flat elliptical disk, which is given by\\

$I=\left(\begin{array}{l}
\sum_im_i(y_i-y_{cm})^2 \hspace{0.7cm} -\sum_im_i(x_i-x_{cm})(y_i-y_{cm})\\
-\sum_im_i(x_i-x_{cm})(y_i-y_{cm}) \hspace{0.7cm} \sum_im_i(x_i-x_{cm})^2\\
\end{array}\right),\\$

where $x_{cm}$ and $y_{cm}$ are the coordinates of the centre of mass of the droplet. The mass $m_i$ of the $i$-th lattice site is related to the local
concentration $\phi_i$ as follows:
\begin{equation}
m_i=1-\frac{(\phi_i-\phi_{min})}{\phi_{max}-\phi_{min}},
\end{equation}
where $\phi_{min}$ and $\phi_{max}$ are the minimum and the maximum values of $\phi$ in the system. In particular $\phi_{max}=2$ in the nematic phase 
and $\phi_{min}=0$ in the isotropic phase. With this definition, $m_i\simeq 1$ in the isotropic phase and $m_i\simeq 0$ in the liquid crystal phase. 
In addition, for numerical stability we have imposed a cutoff on the mass values, namely we have set $m_i=0$ when $m_i<0.04$. 
By diagonalizing the matrix $I$, one gets the principal axes of inertia, which coincide with the $x$ and $y$ axes in an unsheared system.
The principal moments of inertia are $1/\sqrt{I_x}$ and $1/\sqrt{I_y}$, where $I_x=Mb^2/4$ and $I_y=Ma^2/4$ are the eigenvalues of I,
$M=\sum_im_i$, and $a$ and $b$ are the semiaxes. An estimation of the perimeter can be thus obtained through the Ramanujan formula~\cite{villa}
\begin{equation}
\Sigma=\pi\left[(a+b)+\frac{3(a-b)^2}{10(a+b)+\sqrt{a^2+14ab+b^2}}\right].
\end{equation} 

Lastly, the angle that the major axis forms with the direction of the shear flow can be calculated from the eigenvectors of the matrix.


\begin{thebibliography}{99}
\bibitem{musevic_11} I. Musevic, M. Skarabot, M. Humar, {\it J. Phys.: Condens. Matter}, {\bf 23}, 284112 (2011).
\bibitem{paul} P. Poulin, H. Stark, T. Lubensky, and D. Weitz, {\it Science} {\bf 275}, 1770 (1997).
\bibitem{paul2} P. Poulin, D. A. Weitz, {\it Phys. Rev. E} {\bf 57}, 1 (1998).
\bibitem{Yada_04} M. Yada, J. Yamamoto, and H. Yokoyama, {\it Phys. Rev. Lett.} {\bf 92}, 185501 (2004).
\bibitem{Smalyukh_05} I. I. Smalyukh, O. D. Lavrentovich, A. N. Kuzmin, A. V.  Kachynski, and P. N. Prasad,
{\it Phys. Rev. Lett.} {\bf 95}, 157801 (2005).
\bibitem{Nazarenko_01} V. G. Nazarenko, A. B. Nych, and B. I. Lev, {\it Phys. Rev. Lett.} {\bf 87}, 075504 (2001).
\bibitem{Mondain_99} O. Mondain-Monval, J. C. Dedieu, T. Gulik-Krzywicki, and P.  Poulin, {\it Eur. Phys. J. B} {\bf 12}, 167 (1999).
\bibitem{Gu_00} Y. Gu and N. L. Abbott, {\it Phys. Rev. Lett.} {\bf 85}, 4719 (2000).
\bibitem{Fukuda_01} J. Fukuda and H. Yokoyama, {\it Eur. Phys. J. E } {\bf 4}, 389 (2001).
\bibitem{Andrienko_01} D. Andrienko, G. Germano, and M. P. Allen, {\it Phys. Rev. E} {\bf 63}, 041701 (2001).
\bibitem{Allen_05} M. Allen, {\it Comput. Phys. Commun.} {\bf 169}, 433 (2005).
\bibitem{Billeter_00} J. L. Billeter and R. A. Pelcovits, {\it Phys. Rev. E} {\bf 62}, 711 (2000).
\bibitem{gettelfinger} B. T. Gettelfinger, J. A. Moreno-Razo, G. M. Koenig Jr., J. P. H.-Ortiz, N. L. Abbott, J. J. de Pablo, {\it Soft Matter} {\bf 6}, 896 (2010).
\bibitem{Grollau_03} S. Grollau, E. B. Kim, O. Guzman, N. Abbott, and J. J. de Pablo, J. Chem. Phys. 119, 2444 2003;
\bibitem{Ruhwandl_97} R. W. Ruhwandl and E. M. Terentjev, {\it Phys. Rev. E} {\bf 56}, 5561 (1997).
\bibitem{Zhou_et_al_JFM_2007} C. Zhou, P. Yoe and J.J. Feng, {\it J. Fluid. Mech.} {\bf 593}, 385 (2007).
\bibitem{Stark_99} H. Stark, {\it Eur. Phys. J. E } {\bf 10}, 311 (1999).
\bibitem{Araki_04} T. Araki and H. Tanaka {\it Phys. Rev. Lett.} {\bf 93}, 015702 (2004).
\bibitem{Fukuda_04} J. Fukuda, H. Stark, M. Yoneya, and H. Yokoyama, {\it J. Phys.: Condens. Matter } {\bf 16}, S1957  (2004).
\bibitem{lishchuk} S. V. Lishchuk, C. M. Care, {\it Phys. Rev. E} {\bf 70}, 011702 (2004).
\bibitem{care} C. M. Care, I. Halliday, K. Good, S. V. Lishchuk, {\it Phys. Rev. E} {\bf 67}, 061703 (2003).
\bibitem{sulaiman} N. Sulaiman, D. Marenduzzo and J. M. Yeomans, {\it Phys. Rev. E} {\bf 74}, 041798 (2006).
\bibitem{beris} A. N. Beris, B, J. Edwards, {\it Thermodynamics of Flowing Systems}, Oxford University Press, Oxford (1994).
\bibitem{degennes} P. G. de Gennes and J. Prost, {\it The Physics of Liquid Crystals, 2nd Ed.}, Clarendon Press, Oxford (1993).
\bibitem{halperin}  P. C. Hohenberg and B. I. Halperin, {\it Rev. Mod. Phys.} {\bf 49}, 435 (1977).
\bibitem{tiribocchi1} A. Tiribocchi, N. Stella, G. Gonnella and A. Lamura, {\it Phys. Rev. E} {\bf 80}, 026701 (2009).
\bibitem{tiribocchi2} A. Tiribocchi, O. Henrich, J. S. Lintuvuori and D. Marenduzzo, {\it Soft Matter} {\bf 10}, 4580 (2014).
\bibitem{prl} A. Tiribocchi, G. Gonnella, D. Marenduzzo, E. Orlandini, F. Salvadore, {\it Phys. Rev. Lett.}, 2011, {\bf 107}, 237803.
\bibitem{cates1}  M. E. Cates, O. Henrich, D. Marenduzzo, K. Stratford, {\it Soft Matter} {\bf 5}, 3791 (2009).
\bibitem{elsen} E. Tjhung, D. Marenduzzo, M. E. Cates, {\it  Proc. Natl. Acad. Sci. USA} {\bf 109}, 12381-12386 (2012).
\bibitem{elsen2} E. Tjhung, A. Tiribocchi, D. Marenduzzo, M. E. Cates, {\it Nature Comm.} {\bf 6}, 5420 (2015).
\bibitem{giulio} G. De Magistris, A. Tiribocchi, C. A. Whitfield, R. J. Hawkins, M. E. Cates and D. Marenduzzo, {\it Soft Matter} {\bf 10}, 7826 (2014).
\bibitem{cognard} J. Cognard, {\it Mol. Cryst. and Liq. Cryst.} Supp. 1, London (1982).
\bibitem{anderson} V. J. Anderson, E. M. Terentjev, S. P. Meeker, J. Crain and W. C. K. Poon, {\it Eur. Phys. J. E} {\bf 4}, 11 (2001).
\bibitem{faetti} S. Faetti, V. Palleschi, {\it Journ. Chem Phys.} {\bf 81}, 6254 (1981).
\bibitem{luben} T. C. Lubensky, D. Pettey, N. Currier, and H. Stark, {\it Phys. Rev. E} {\bf 57}, 610 (1998).
\bibitem{sluckin} N. Schopohl, T. J. Sluckin, {\it Phys. Rev. Lett.} {\bf 60}, 755 (1988).
\bibitem{callan} A. C. Callan Jones, R. A. Pelcovits, V. A. Slavin, S. Zhang, D. H. Laidlaw, G. B. Loriot, {\it Phys. Rev. E} {\bf 74}, 061701 (2006).
\bibitem{taylor} G. I. Taylor, {\it Proc. R. Soc. Lond. Ser. A}, {\bf 146}, 501 (1934).
\bibitem{olmsted} P. D. Olmsted, P. M. Goldbart, {\it Phys. Rev. A} {\bf 46}, 8 (1992).
\bibitem{renardy} Y. Y. Renardy, V. Cristini, {\it Phys. of Fluids} {\bf 13}, 1 (2001).
\bibitem{Wagner} A. J. Wagner and J. M. Yeomans {\it Int. Journ. Mod. Phys. C} {\bf 8}, 773 (1997).
\bibitem{Khullar_et_al_PRL_2007} S. Khullar, C. Zhou and J.J. Feng, {\it Phy. Rev. Lett.} {\bf 99}, 237802 (2007).
\bibitem{villa} M. B. Villarino, {\it Journ. Ineq. Pure and Appl. Math.} {\bf 7}, 1 (2006).

\end{thebibliography}
\end{document}